\documentclass[fleqn,usenatbib]{mnras}

\usepackage{newtxtext,newtxmath}

\usepackage[T1]{fontenc}

\DeclareRobustCommand{\VAN}[3]{#2}
\let\VANthebibliography\thebibliography
\def\thebibliography{\DeclareRobustCommand{\VAN}[3]{##3}\VANthebibliography}

\newcommand{\ud}{\mathrm{d}}

\usepackage{graphicx}	
\usepackage{amsmath}	
\usepackage{tabularx}
\usepackage{caption}  
\usepackage{subcaption}

\title[Resolution Dependence of MHD CCSN simulations]{Resolution Dependence in Magnetohydrodynamic Simulations of Neutrino-Driven Core-Collapse Supernovae}

\author[Varma \& M\"uller]{
Vishnu Varma$^{1}$\thanks{E-mail: 
v.r.vejayan@keele.ac.uk},
Bernhard M\"uller$^{2}$
\\
$^{1}$
{Astrophysics Group, Lennard-Jones Laboratories, Keele University, Keele ST5 5BG, UK}\\
$^{2}$
School of Physics and Astronomy, 10 College Walk, Monash University, Clayton, VIC 3800, Australia\\
}

\date{Accepted XXX. Received YYY; in original form ZZZ}

\pubyear{2015}

\begin{document}
\label{firstpage}
\pagerange{\pageref{firstpage}--\pageref{lastpage}}
\maketitle
\begin{abstract}
We investigate the role of resolution and initial magnetic field strength on core-collapse supernovae in simulations of a non-rotating $13 \mathrm{M_\odot}$ progenitor. Specifically, we study the effect on shock revival, explosion dynamics, and the properties of the compact remnant. We run four models with different numerical grid resolutions with an initial central dipole field strength of $\mathord{\approx}10^{12}\, \mathrm{G}$. Two of those resolutions are also run with a weaker central magnetic field of $\mathord{\approx}10^{10}\, \mathrm{G}$ . The shock revival time for all models is largely independent of resolution and initial magnetic field strength, but we find higher explosion energies when the initial magnetism is stronger and at higher resolutions. We find that models with strong magnetic fields have lower neutrino luminosity and energies, due to a proto-neutron star (PNS) that is deformed by the strong magnetic fields. At higher resolutions, magnetic fields are amplified more efficiently in the gain region and in the PNS via the small-scale dynamo. Although the strong magnetic fields do not directly drive the explosion, they have a subsidiary impact on the explosion mechanism and compensate for the reduced neutrino heating. Stronger magnetic energies in the PNS also affect energy and angular momentum redistribution, leading to more extended and vigorous PNS convection zones at higher resolutions. 
\end{abstract}




\section{Introduction}
\label{sec:Intro} 
Core-collapse supernovae (CCSNe) are the explosions of massive stars (M $\geq 8\mathrm{M}_\odot$), signalling the end of their lives. Among the proposed explosion mechanisms for these CCSNe events, the neutrino-driven mechanism initially suggested by \citet{Colgate1966} has been the most extensively explored over the last few decades. In this mechanism, shock revival is accomplished because of the increase in the post-shock pressure from the partial reabsorption of neutrinos that stream out from the young proto-neutron star (PNS), which is further aided by multi-dimensional effects \citep{Herant1995, Burrows1995, Blondin2003, Janka2007} 

Improvements in computational power and numerical methods have recently allowed for in-depth studies of 3D neutrino-driven CCSN, including more detailed physics such as multi-group neutrino transport, magnetic fields, and  general relativity, however, even the most sophisticated simulations still need to make approximations
(for recent reviews, see \citealt{Muller_2016review,Mueller_LRCA,Mezzacappa_LRCA, Burrows2020, Janka2025}).

Rapidly rotating, strongly magnetic supernovae (i.e., magnetorotational supernovae) have also received significant attention over the years, 
in particular as a scenario for rare hypernovae and gamma-ray bursts
\citep[e.g.][]{Leblanc1970, Bisnovatyi-Kogan1976, Mueller1979, Burrows2007,Obergaulinger2009, Sawai2013, Masada2015, Moesta2015, Bugli2020, Obergaulinger2021, Powell2023}, and as a possible site for r-process nucleosynthesis \citep{Moesta2018, Reichert2021, Zha2024}. This mechanism relies on the amplification of magnetic fields by differential rotation in the CCSN core, e.g., by the magnetorotational instability (MRI; \citealt{Balbus1991, Akiyama2003})
or an $\alpha$-$\Omega$-dynamo in the PNS \citep{duncan_92,Thompson1993,Raynaud2020}. Strong magnetic stresses can then drive the explosion, often in the form of collimated jets.

Observational evidence of a large population of neutron stars with strong magnetic fields \citep[magnetars,][]{popov_12}, however, suggests that such strong
fields are a more common
phenomenon in core-collapse
supernovae, and it is conceivable that they also have some impact on supernova explosion dynamics. The study of magnetic fields in the 
explosions of slowly rotating or non-rotating massive stars (see recent review \citealt{Mueller2024}), however, has so far received less attention \citep{Obergaulinger2014, Matsumoto2020, MullerVarma2020, Varma2023, Matsumoto2024, Nakamura2024, Sykes2024} than the more spectacular scenario of magnetorational explosions. 
By contrast,
magnetic field amplification cannot feed on a large reservoir of rotational energy in this case, fields of significant strength may still
be generated by a small-scale dynamo \citep{MullerVarma2020} or by the standing accretion shock instability \citep{Endeve2012}.
It has also been theorised that strong pre-supernova fields of order $10^{12}\, \mathrm{G}$ might already be present in some supernova progenitors as fossil fields from a stellar merger \citep{ferrario_05,ferrario_06,SchneiderNature,Schneider2020}. The strongly magnetised regime we explore here is further motivated by recent advances in 3D magnetohydrodynamic shell burning simulation of \citep{VarmaMuller2021, Leidi2023, VarmaMuller2023}, which suggests that shell convection in the final seconds prior to core-collapse can generate rather strong magnetic fields of $10^{10\texttt{-}11}\,\mathrm{G}$ in the oxygen shell. Consequently, the impact of magnetic fields in core-collapse supernovae of
non-rotating or slowly rotating progenitors has
received growing interest in recent years, with several 3D simulation studies investigating their effect
on shock revival, explosion dynamics and compact remnants birth properties \citep{MullerVarma2020,Varma2023, Matsumoto2022, Matsumoto2024, Sykes2024}.


The complexity of CCSN codes and the disparities in numerical methodology present a challenge for drawing robust conclusions for such studies.
For hydrodynamic simulations of neutrino-driven
supernovae, extensive code comparisons \citep{Liebendorfer2005, Mueller2010, OConnor2018, Just2018, Cabezon2018} and resolution studies \citep[e.g.,][]{Abdikamalov2014,Handy2013,Hanke2012,Couch2014a,Radice2015,Nagakura2019,Melson2020}
have been performed to gauge the robustness of  of the results and uncertainties related to different numerical recipes and input physics.


However, such studies have yet to be performed for simulations of 3D MHD neutrino-driven supernovae. Even for the better explored magnetorotational mechanism, very few resolution studies and code comparisons have been performed. These studies have been limited to 2D global simulations \citep[e.g.,][]{Sawai2015,Varma2021} or more detailed local studies performed to understand the growth of the magnetorotational instability \citep{Obergaulinger2009,Masada2015}. Even without rotation to amplify the field, the highly turbulent environment in the CCSN gain region and in the forming PNS can still grow the magnetic energies in these regions through a small-scale dynamo (SSD, \citep{Brandenburg2005, Tobias2021,Rempel2023}). Studies of this dynamo mechanism \citep{Schekochihin2007} have shown a clear resolution dependence of the SSD growth rate.

While a difference in the growth rate of the magnetic field may not qualitatively impact the long-term evolution in steady-state stellar convection \citep{Leidi2023}, a higher resolution, and hence more resolved turbulence, has been shown to change how energy and angular momentum can be redistributed \citep{Hotta2021, Mori2023}. Better resolved 3D hydrodynamic turbulence in the gain region has already been shown to impact the neutrino-driven mechanism \citep{Nagakura2019, Melson2020}, so the impact of the resulting SSD that inevitably ensues is important to study.

In this study, we present a resolution study of a strongly magnetised, non-rotating neutrino-driven supernova models in 3D, with four different grid resolutions. For two of these resolutions, we also compare to an equivalent weakly magnetised model to better understand how the strength of the magnetic field aids the neutrino-driven mechanism. 

Our paper is structured as follows: In Section~\ref{sec:Initial}, we describe the progenitor model as well as the initial conditions of our simulations. This is followed by a description of the numerical methods of the \textsc{CoCoNuT-FMT} code in Section~\ref{sec:methods}. The results of the simulations are presented in Section~\ref{sec:results}, where we first look at the large-scale explosion dynamics. We then analyse the role of resolution and magnetic fields on the neutrino properties, followed by an analysis of their impact on the gain region dynamics and the resulting PNS properties. We summarise our results and discuss their implications in Section~\ref{sec:conclusion}.

\section{Progenitor Model and Initial Conditions}
\label{sec:Initial} 

\begin{figure}
 \includegraphics[width=\columnwidth]{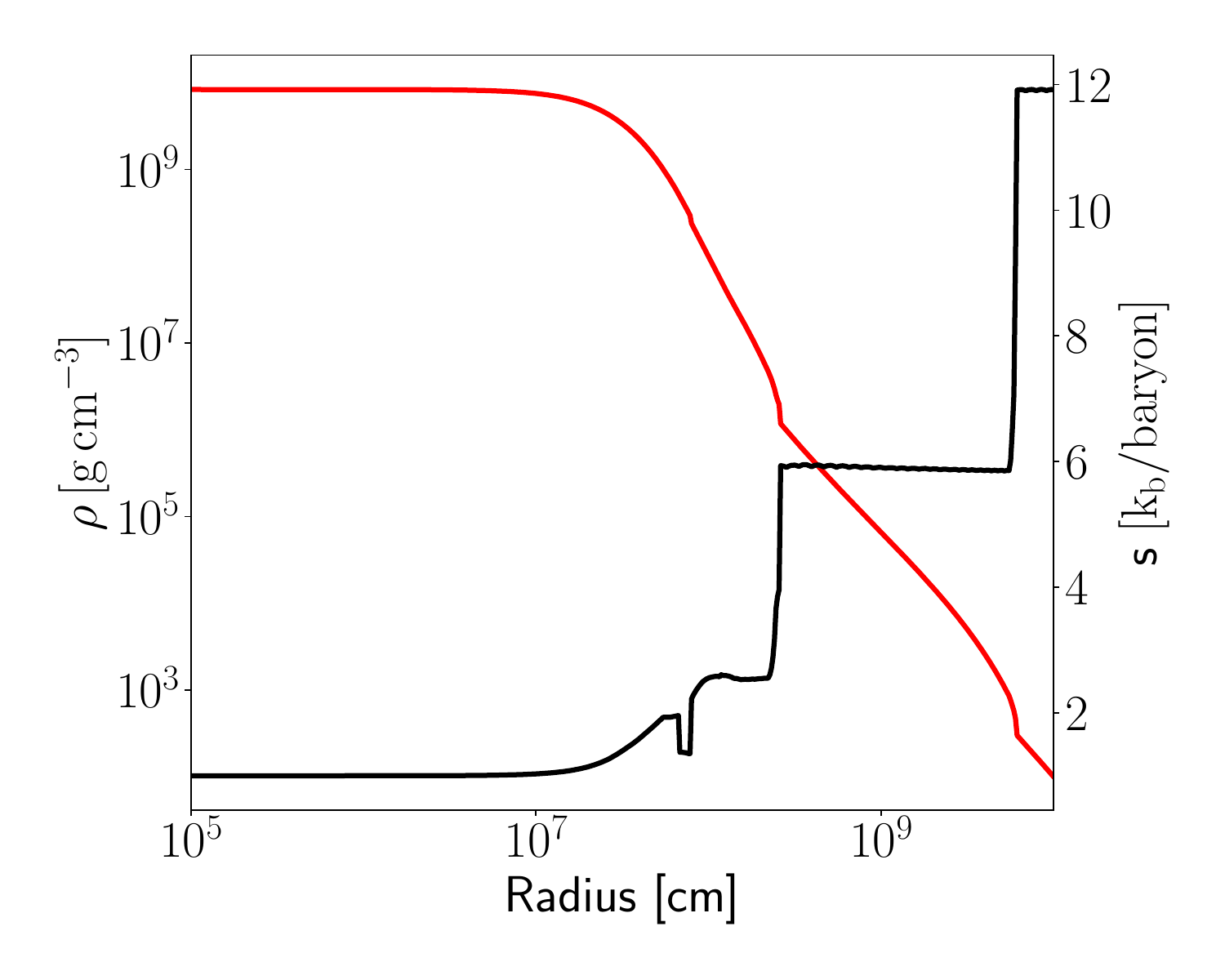}
 \caption{\normalsize Initial density (red) and entropy (black) profiles mapped from \textsc{MESA} to \textsc{CoCoNuT-FMT}.}
 \label{fig:Initial}  
\end{figure}

We simulate the core collapse of a non-rotating, magnetised massive star of 13$\mathrm{M_\odot}$. This progenitor is published in \citet{Whitehead2026} (model labelled 13SH21 in the publication), where it was run using the \textsc{MESA} 1D stellar evolution code \citep{Paxton2011, Paxton2019} with a metallicity of $Z = 0.001$, using convective boundary mixing parameters as described by \citet{Scott2021}. This progenitor was then mapped to the \textsc{CoCoNuT-FMT} 3D magnetohydrodynamic supernova code at the onset of core collapse when the infall velocity has reached 1000\,$\mathrm{km\, s^{-1}}$. The entropy and density profiles at the time of collapse are shown in Figure \ref{fig:Initial}.

In our 3D simulations, we impose a dipolar magnetic field with two different field strengths. Strong field runs have $B_\mathrm{pol,tor} = 10^{12}\,\mathrm{G}$ and weak field cases have $B_\mathrm{pol,tor} = 10^{10}\,\mathrm{G}$ at the centre of the star.
The magnetic field geometry is defined using the vector potential $\mathbf{A}$ from \citet{Suwa2007, Obergaulinger2018, Varma2021},
\begin{align}
\label{eq:magini}
    \left(A^{r}, A^{\theta}, A^{\phi}\right)=
\frac{1}{2\left(r^{3}+r_{0}^{3}\right)}
    \left(B_\mathrm{tor} r_{0}^{3} r \cos \theta, 0, B_\mathrm{pol} r_{0}^{3} r \sin \theta\right),
\end{align}
where the radius parameter is set to $r_0=10^8\,\mathrm{cm}$.

Recent magnetoconvection simulations of the final phases of shell convection in supernova progenitors \citep{VarmaMuller2021, Leidi2023, VarmaMuller2023} indicate that the pre-supernova field geometry likely develops significant small-scale structure. Here, we opt for a simple dipole geometry to reduce the complication of our comparison. It should also be noted that we do not possess a full understanding of the global magnetic field geometry in the star, so any increase in complication to this geometry is ultimately a free parameter.

For this resolution analysis, we used four different resolution choices for the $B= 10^{12}\,\mathrm{G}$ models and two choices for the $B = 10^{10}\,\mathrm{G}$ models to study how resolution affects the initial magnetic field strengths. 
The models labelled "MRes" are run with the resolution ($N_r$ $\times$ $N_\theta$ $\times$ $N_\phi$) = (550 $\times$ 128 $\times$ 256). The LRes model uses half the angular grid number of MRes, vLRes halves the grid count of all 3 components, while HRes uses twice the angular number of MRes. The two models with weaker magnetic fields (i.e. $B = 10^{10}\,\mathrm{G}$) are called MRes10 and vLRes10 respectively. 
The full list of simulated models, their resolution choices and initial magnetic field strengths are listed in Table \ref{tab:resolution}.

\begin{table*}
    \centering
    \begin{tabular}{ccclc} \hline \hline
         \textbf{Model}&  \textbf{Resolution} ($N_r$ $\times$ $N_\theta$ $\times$ $N_\phi$)&  \textbf{Angular resolution}&{$\mathbf{\Delta V}$}& \textbf{Initial Field [G]}
\\ \hline 
         vLRes&  275$\times$64$\times$128&  $2.8^\circ$&69.12$\mathrm{km^3}$& $10^{12}$
\\ \hline 
         LRes&  550$\times$64$\times$128&  $2.8^\circ$&34.56$\mathrm{km^3}$& $10^{12}$
\\ \hline 
         MRes&  550$\times$128$\times$256& $1.4^\circ$&8.64$\mathrm{km^3}$& $10^{12}$
\\ \hline 
         HRes&  550$\times$256$\times$512&  $0.7^\circ$&2.16$\mathrm{km^3}$& $10^{12}$
\\ \hline 
         vLRes10&  275$\times$64$\times$128&  $2.8^\circ$&69.12$\mathrm{km^3}$& $10^{10}$
\\ \hline 
         MRes10&  550$\times$128$\times$256&  $1.4^\circ$&8.64$\mathrm{km^3}$& $10^{10}$
\\\hline\hline
    \end{tabular}
    \caption{A list of the initial conditions of all models presented in this paper. Each model uses the same 13$\mathrm{M_\odot}$ progenitor but is run with a different set of resolution and initial magnetic field conditions. $\Delta V$ is the volume of a grid cell located at $r=100\,\mathrm{km}$ at the equator for each model.
    }
    \label{tab:resolution}
\end{table*}

\section{Numerical Methods}
\label{sec:methods} 
We employ the Newtonian magnetohydrodynamic (MHD) version of the  \textsc{CoCoNuT-FMT} code as described in detail in \citet{MullerVarma2020, Varma2021} and \citet{Varma2023}. 

The MHD equations are solved in spherical
polar coordinates using the HLLC (Harten-Lax-van Leer-Contact) Riemann solver \citep{Gurski2004, Miyoshi2005}, piecewise parabolic reconstruction
\citep{Colella1984}, and mesh coarsening to reduce numerical dissipation near the grid axis \citep{Muller2019a, Varma2023}. The divergence-free condition $\nabla\cdot\mathbf{B} = 0$ is maintained using a modified version
of the original hyperbolic divergence cleaning scheme of \citet{Dedner2002} and follows ideas of \citet{Tricco2016} to maintain energy conservation. Compared to the original cleaning method, we rescale
the Lagrange multiplier $\psi$ to
$\hat{\psi}=\psi/c_\mathrm{h}$, where $c_\mathrm{h}$ is the hyperbolic cleaning speed. Details of this approach are discussed in the Appendix of \citet{VarmaMuller2021}.
The extended system of MHD equations for the density $\rho$, velocity $\mathbf{v}$, magnetic field $\mathbf{B}$, the total energy density $\hat{e}$, mass
fractions $X_i$, and the rescaled Lagrange multiplier $\hat{\psi}$ reads,
\begin{eqnarray}
\partial_t \rho
+\nabla \cdot \rho \mathbf{v}
&=&
0,
\\
\partial_t (\rho \mathbf v)
+\nabla \cdot \left(\rho \mathbf{v}\mathbf{v}-
\frac{\mathbf{B} \mathbf{B}}{4\pi}
+P_\mathrm{t}\mathcal{I}
\right)
&=&
\rho \mathbf{g}
-
\frac{(\nabla \cdot\mathbf{B}) \mathbf{B}}{4\pi}
,
\\
\partial_t {\hat{e}}+
\nabla \cdot 
\left[(e+P_\mathrm{t})\mathbf{v}
-\frac{\mathbf{B} (\mathbf{v}\cdot\mathbf{B})
{-c_\mathrm{h} \hat{\psi} \mathbf{B}}}{4\pi}
\right]
&=&
\rho \mathbf{g}\cdot \mathbf{v}
,
\\
\partial_t \mathbf{B} +\nabla \cdot (\mathbf{v}\mathbf{B}-\mathbf{B}\mathbf{v})
+\nabla  (c_\mathrm{h} \hat{\psi})
&=&0,
\\
\partial_t \hat{\psi}
+c_\mathrm{h} \nabla \cdot \mathbf{B}
&=&-\hat{\psi}/\tau
,
\end{eqnarray}
where $\mathbf{g}$ is the gravitational acceleration, $P_\mathrm{t}$ is the total pressure, $\mathcal{I}$ is the Kronecker tensor and $\tau$ is the damping timescale for divergence cleaning. This
system conserves the volume integral of a
modified total energy density $\hat{e}$,
which also contains the cleaning field $\hat{\psi}$,
\begin{equation}
{\hat{e}}
=\rho \left(\epsilon+\frac{v^2}{2}\right)+\frac{B^2+\hat{\psi}^2}{8\pi},
\end{equation}
where $\epsilon$ is the mass-specific internal energy. 

Neutrinos are treated using the \textsc{FMT} (fast multi-group transport)
scheme of \citet{Mueller2015}, which solves the energy-dependent
zeroth moment equation for three neutrino species
in the stationary approximation using a closure
obtained from a two-stream solution of the Boltzmann equation. The models employ an exponential
grid in energy space with $21$ zones from $4 \, \mathrm{MeV}$
to $240 \, \mathrm{MeV}$. The simulations use the SFHo equation of state of \citet{Steiner2013} at high densities, and an ideal gas consisting of photons, electrons, positions, and
non-relativistic nucleons and nuclei, in conjunction with an NSE solver above $5 \, \mathrm{GK}$  and
a flashing treatment at lower temperatures \citep{Rampp2000}.

\begin{figure*}
\centering
    \subfloat{\includegraphics[width=0.5\linewidth]{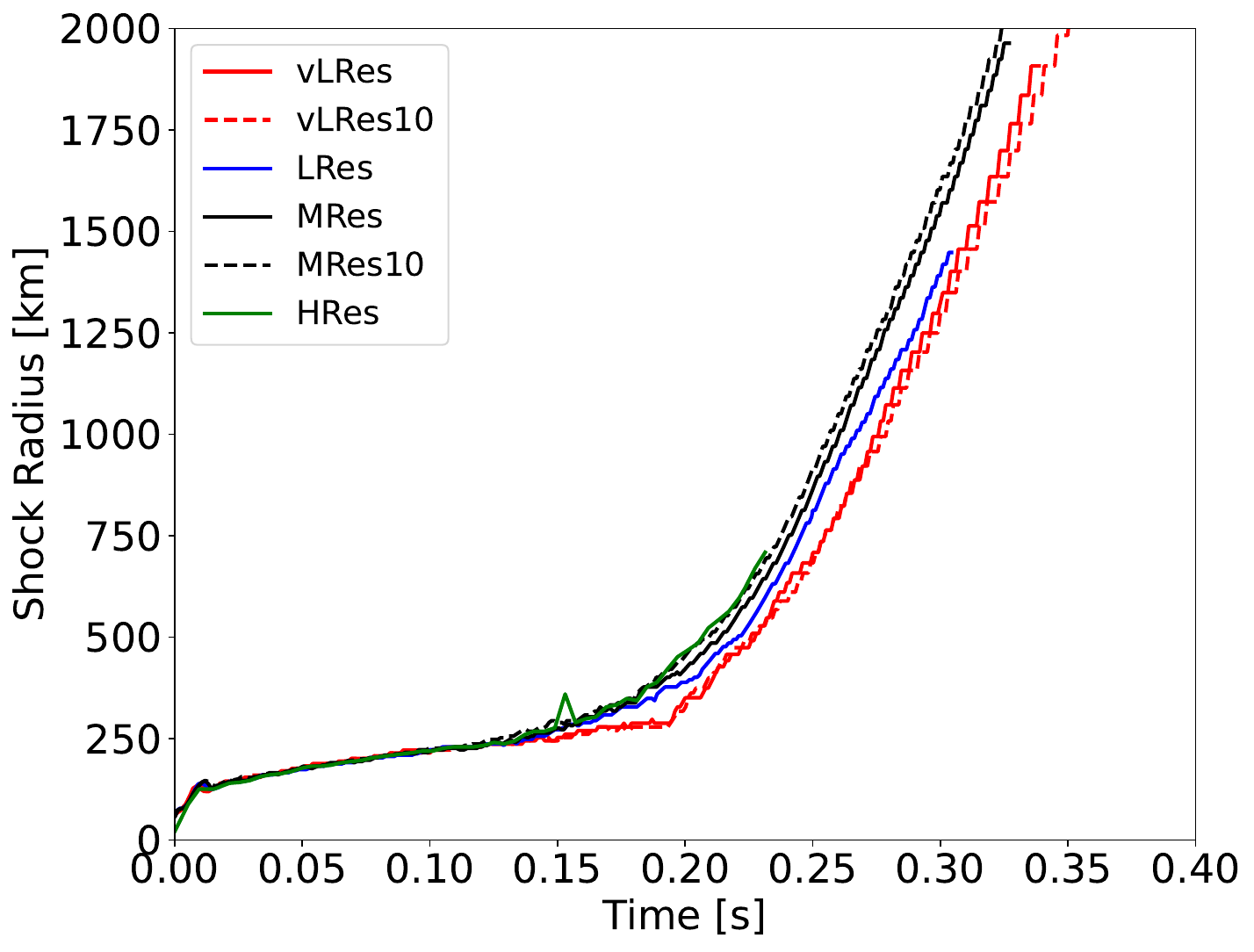}}
\hfil
    \subfloat{\includegraphics[width=0.5\linewidth]{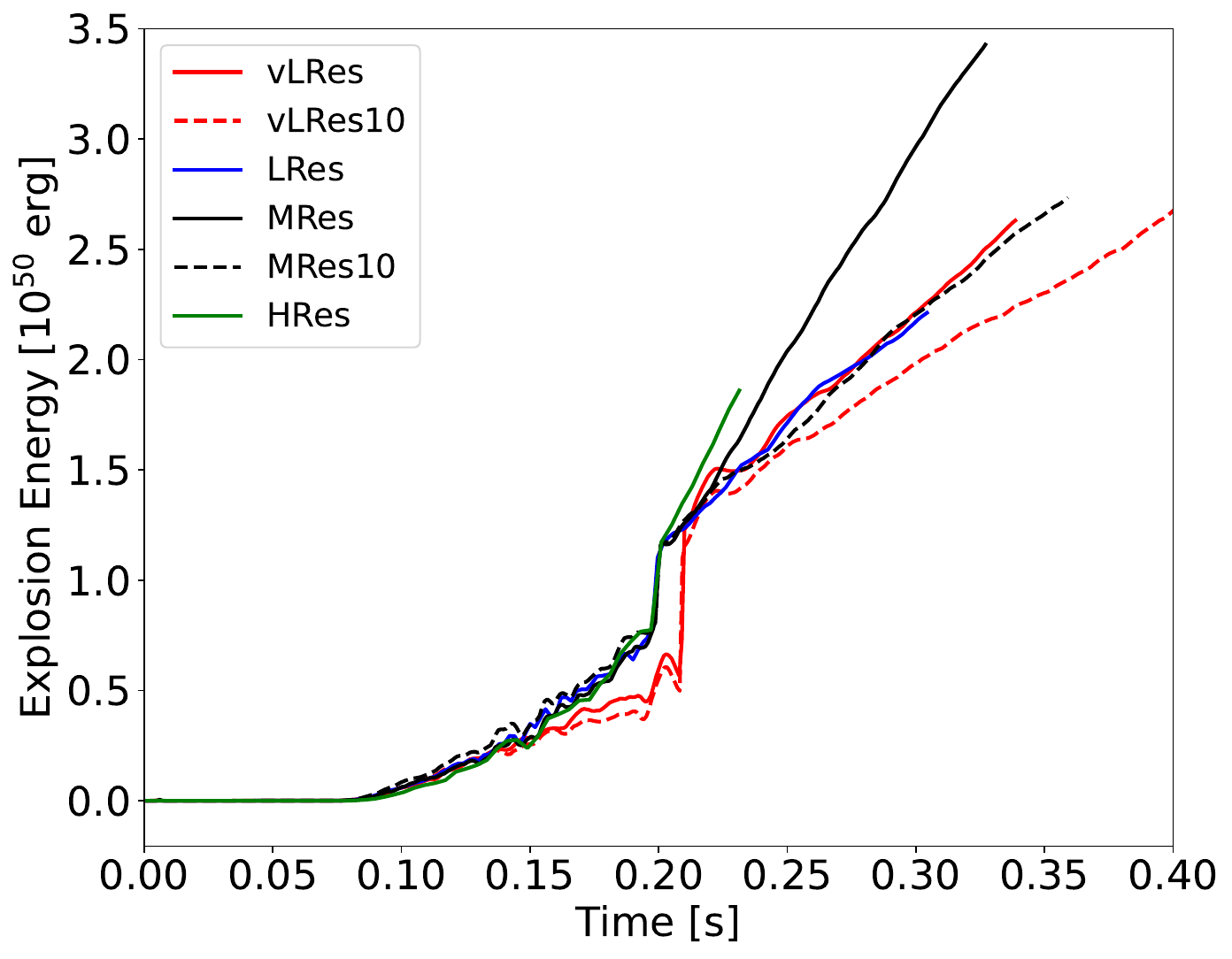}}
\caption{\normalsize The evolution of the mean shock radius (left) and diagnostic explosion energy (right) for the entire set of simulations. }
    \label{fig:Exp_shock}
\end{figure*}

\section{Results}
\label{sec:results} 

To better understand the importance of resolution on the magnetic neutrino-driven mechanism, we simulate six versions of a 13$\mathrm{M_\odot}$ progenitor, varying only the resolution and initial magnetic field strength as noted in Table~\ref{tab:resolution}. 

\subsection{Explosion Dynamics and Morphology}
\label{subsec:dynamics} 

We begin our analysis with a comparison of the bulk dynamics of our suite of models. Figure~\ref{fig:Exp_shock} shows the angle-averaged shock radius (left) along with the diagnostic explosion energy $E_\mathrm{expl}$ \citep{Buras2006} for all models (right). Here, the diagnostic explosion energy is defined as an integral over the region that is nominally unbound,

\begin{equation}
E_\mathrm{expl} = \int\limits_{e_\mathrm{tot}>0} \rho e_\mathrm{tot}\,\mathrm{d}V ,
\end{equation}
and $e_\mathrm{tot}$ is the total energy density, i.e., the sum of the internal, kinetic, gravitational, and magnetic energy density.

\begin{figure*}
\centering
    \begin{subfigure}{0.4\linewidth}
        \includegraphics[width=\linewidth]{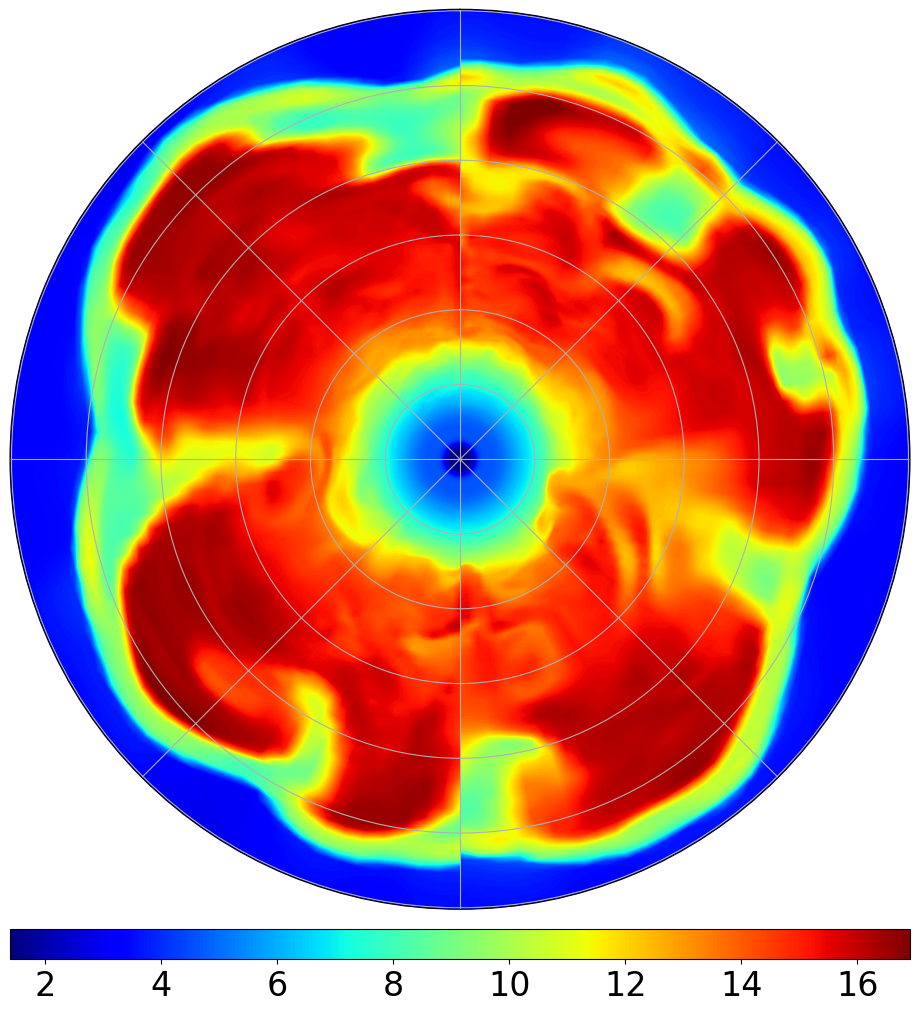}
    \end{subfigure}
\hfil
    \begin{subfigure}{0.4\linewidth}
        \includegraphics[width=\linewidth]{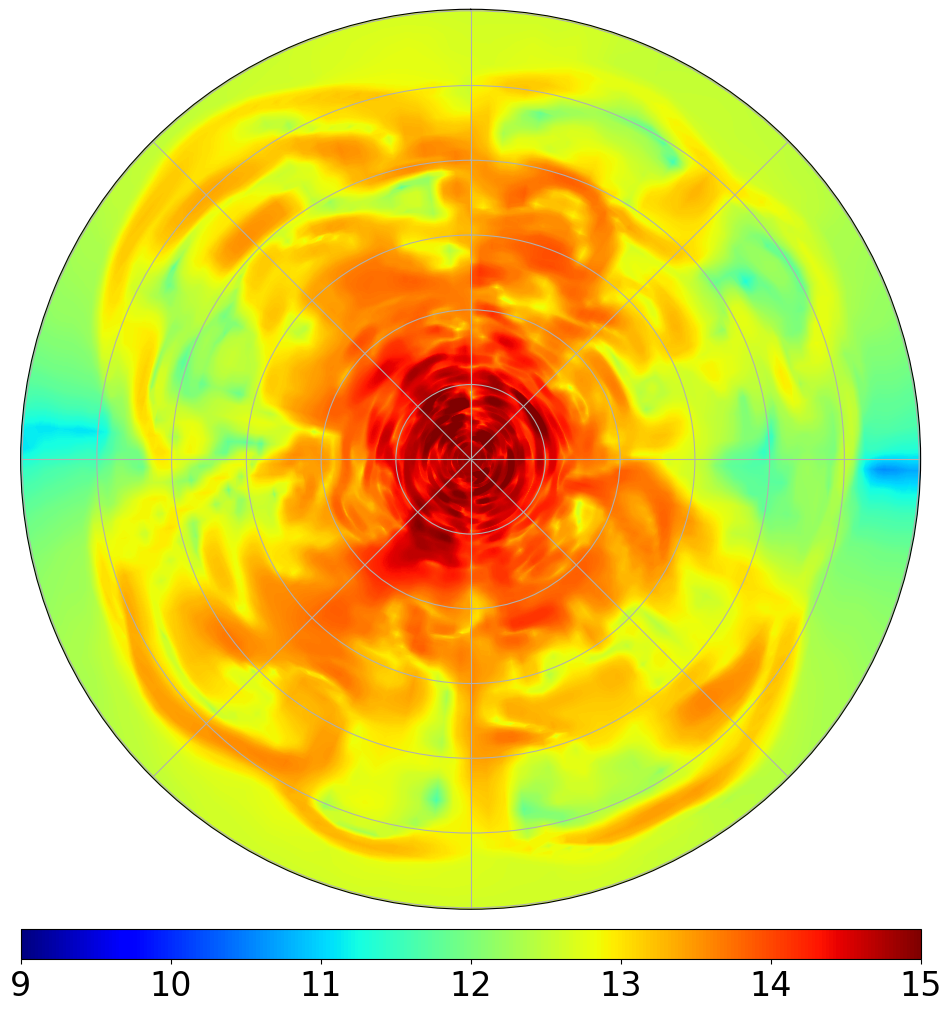}
    \end{subfigure}

    \begin{subfigure}{0.4\linewidth}
        \includegraphics[width=\linewidth]{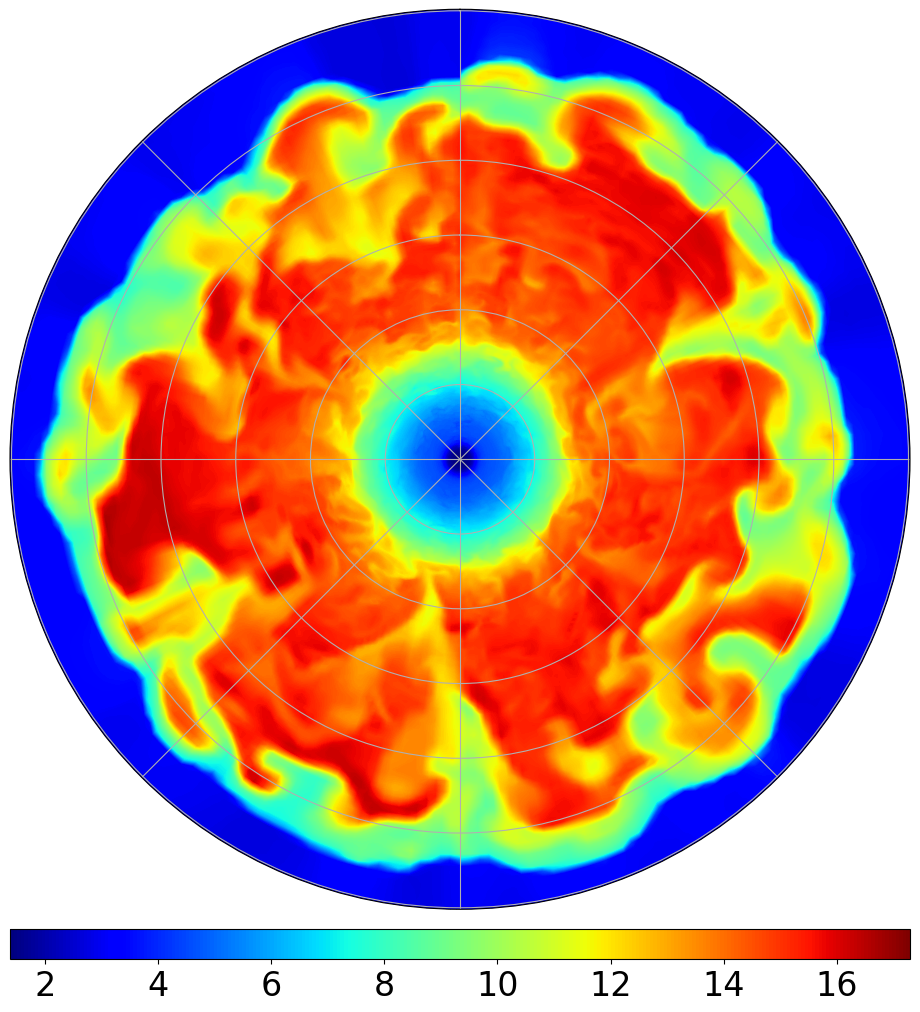}
    \end{subfigure}
\hfil
    \begin{subfigure}{0.4\linewidth}
        \includegraphics[width=\linewidth]{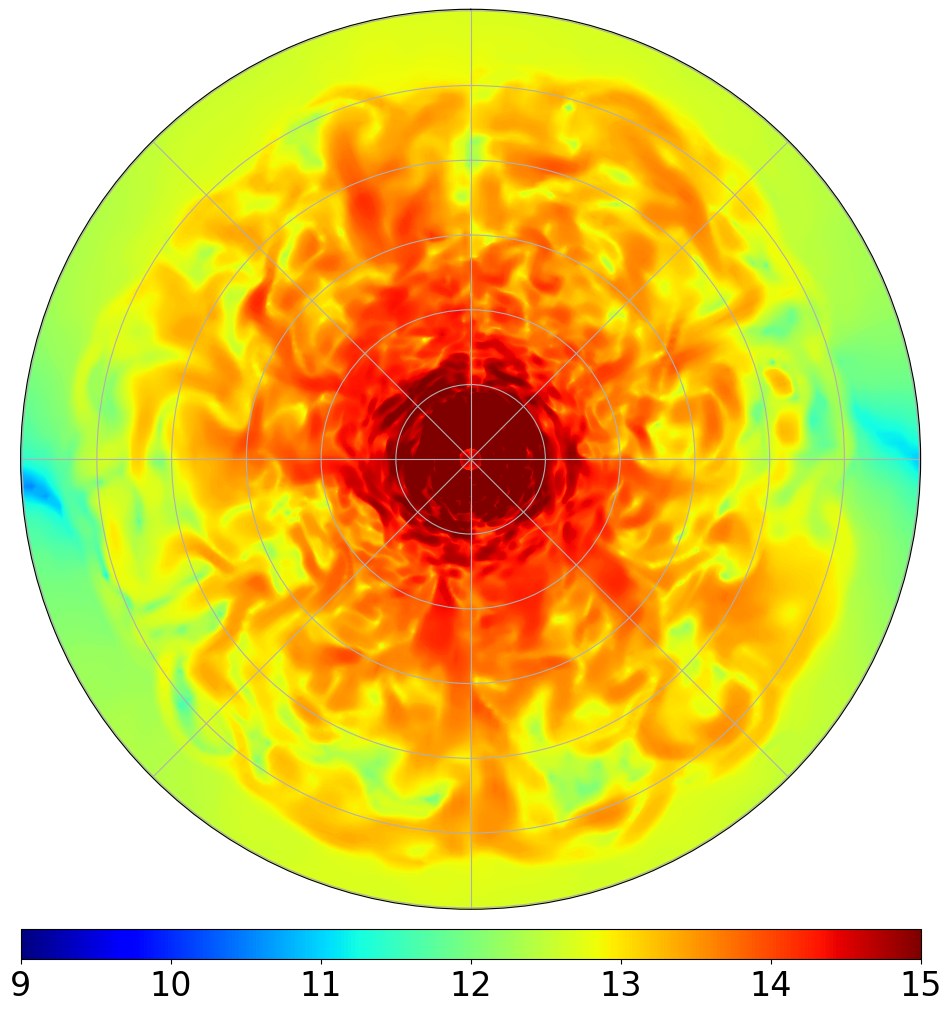}
    \end{subfigure}

    \begin{subfigure}{0.4\linewidth}
        \includegraphics[width=\linewidth]{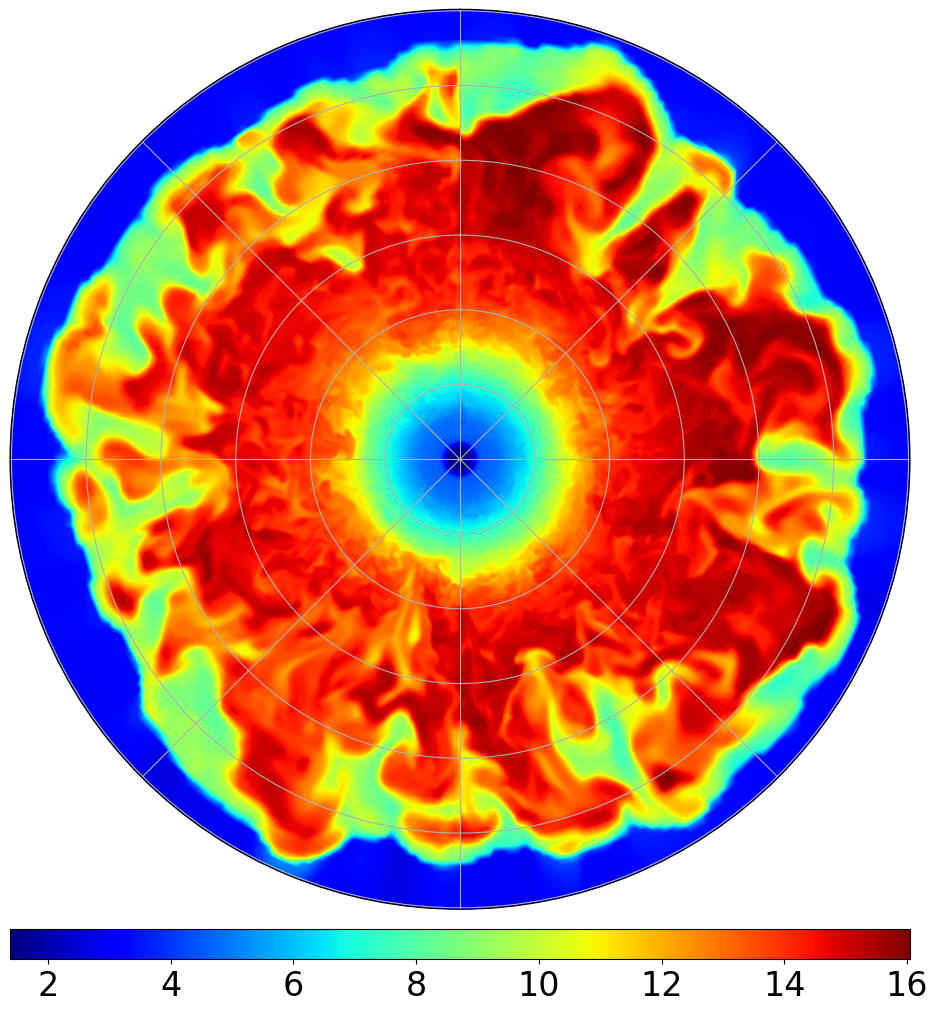}
    \end{subfigure}
\hfil
    \begin{subfigure}{0.4\linewidth}
        \includegraphics[width=\linewidth]{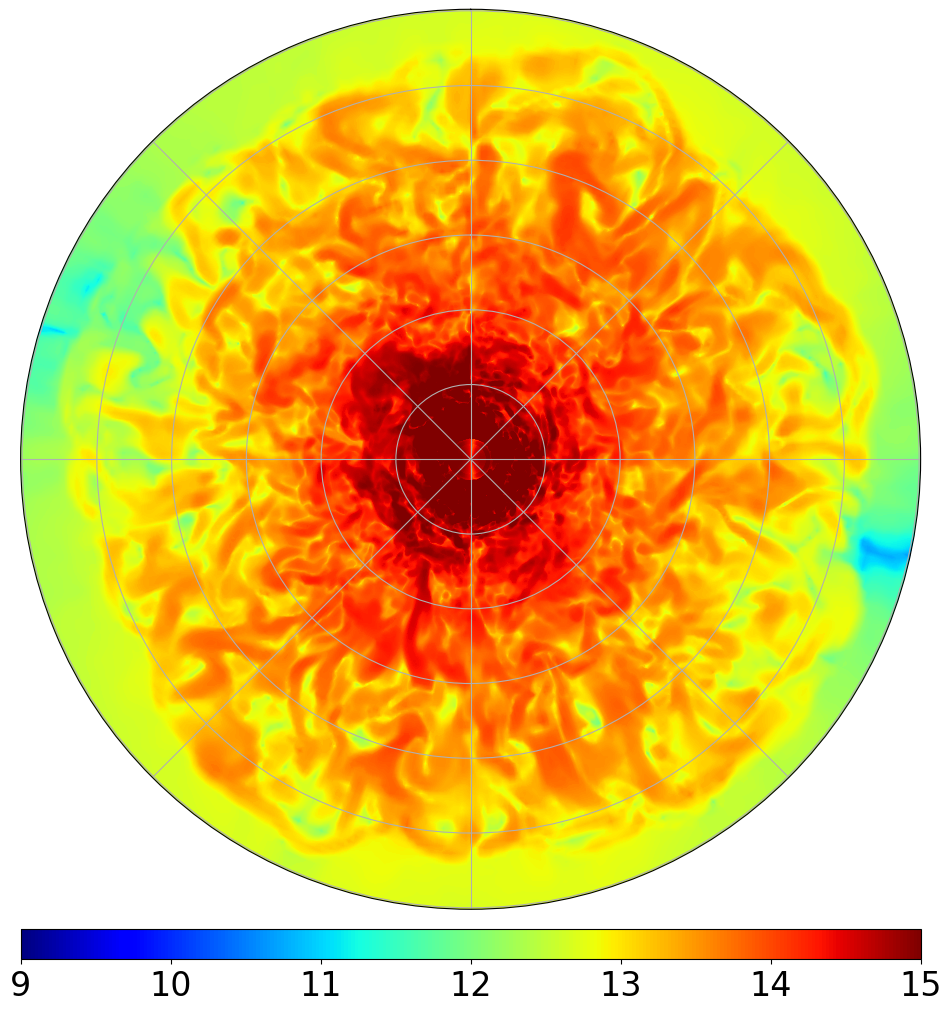}
    \end{subfigure}
\caption{Entropy (left) and root mean squared magnetic field strength (right) on meridional slices for models LRes, MRes and HRes (from top to bottom) models at 0.17\,s post-bounce up to a radius of 300\,km. The entropy colourbars are in units $\mathrm{k_b/baryon}$, and the decadic logarithm of the magnetic field strength in $G$ is plotted  from $10^9$ - $10^{15}$\,G}.
    \label{fig:2D_Slice}
\end{figure*}

We find that the shock radii of all models are initially very similar but start to deviate between 0.15\,s and 0.20\,s post-bounce, where shock revival occurs for these models. Similar to the hydrodynamic models of \citet{Nagakura2019} and \citet{Melson2020}, after revival, the shock expansion appears to be correlated to the resolution, with the highest resolution models expanding the fastest. The highest and lowest resolution models have up to a 38\% difference in shock radii. We note that, aside from small stochastic differences, the average shock radius is largely independent of the initial magnetic field strength. This is shown by comparing models of the same resolution but different initial magnetic field strengths, vLRes to vLRes10 and MRes to MRes10.  
Before shock revival, we do not see any clear difference between the shock expansion in all models, regardless of the resolution. This differs from the hydrodynamic resolution study of \citet{Nagakura2019}, where the increased turbulent pressure leads to a 10-20\,km increase in the stalled shock radius for their high-resolution model. We do not see such an additional effect of turbulence in our MHD models because turbulent kinetic energy is being efficiently converted into magnetic energy, leaving the turbulent pressure reasonably similar between the models. The increase in turbulent energy is instead seen as an increase in magnetic energy in the gain region. We thus do not find any clear signs of the higher resolution models imparting more turbulent pressure.
A visual inspection of Figure~\ref{fig:2D_Slice} shows that increasing resolution leads to progressively finer structure in both the entropy and magnetic field distributions. Entropy plumes become increasingly fragmented, with additional instabilities emerging along plume boundaries. The magnetic field morphology evolves from relatively smooth structures at low resolution to highly intermittent thin filaments at high resolution. These features are consistent with the development of a turbulent cascade and reduced numerical dissipation. Despite the increasing small-scale complexity, the overall large-scale morphology of the convective region remains broadly similar across models. We discuss the dynamics of the gain region in more detail in Section~\ref{subsec:Gain}.

While there is a general trend of increasing explosion energy with increasing resolution in Figure~\ref{fig:Exp_shock}, the increase is less monotonic. From $\approx$0.15\,s the vLRes and vLRes10 models have up to a 25\% lower explosion energy than the other models. This is due to the higher numerical diffusivity that prevents the growth of both turbulent kinetic and magnetic energy in the gain region. This phenomenon will be discussed further below. After $\approx$0.20\,s post-bounce, the vLRes and LRes models are very similar. Interestingly, we find that, unlike the shock radii, the explosion energies are correlated with the initial magnetic field strengths. After runaway expansion occurs for all models, the stronger magnetic field models (vLRes and MRes) have explosion energies that grow faster than their weakly magnetic counterparts (vLRes10 and MRes10). We find that variation in resolution and initial field strength has a similar effect on shock propagation and explosion energy for these models. In particular, the explosion energy of model MRes10 evolves very similarly to model LRes with lower resolution and stronger initial magnetisation. We find a similar phenomenon when analysing the neutrino energies and luminosities in Section~\ref{subsec:Neutrino}. 

We note that all models show a sharp increase in their diagnostic explosion energies around 0.19\,s (and $\approx$0.21\,s for vLRes and vLRes10). This corresponds to the accretion of the density drop at the Si/O interface (see Figure~\ref{fig:Initial}) through the expanding shock. The accretion of the interface has been shown to trigger runaway explosion in some simulations \citep[e.g.,][]{Suwa2016, Summa2016, Vartanyan2018}; however, since our models explode early, the interface is only accreted while the explosion is already in progress. This causes a decrease in ram pressure, and a corresponding increase in the amount of material that is unbound. The Si/O interface is accreted slightly later in models vLRes and vLRes10 due to the slower rate of shock expansion (Figure~\ref{fig:Exp_shock}, left).

\begin{figure}
 \includegraphics[width=\columnwidth]{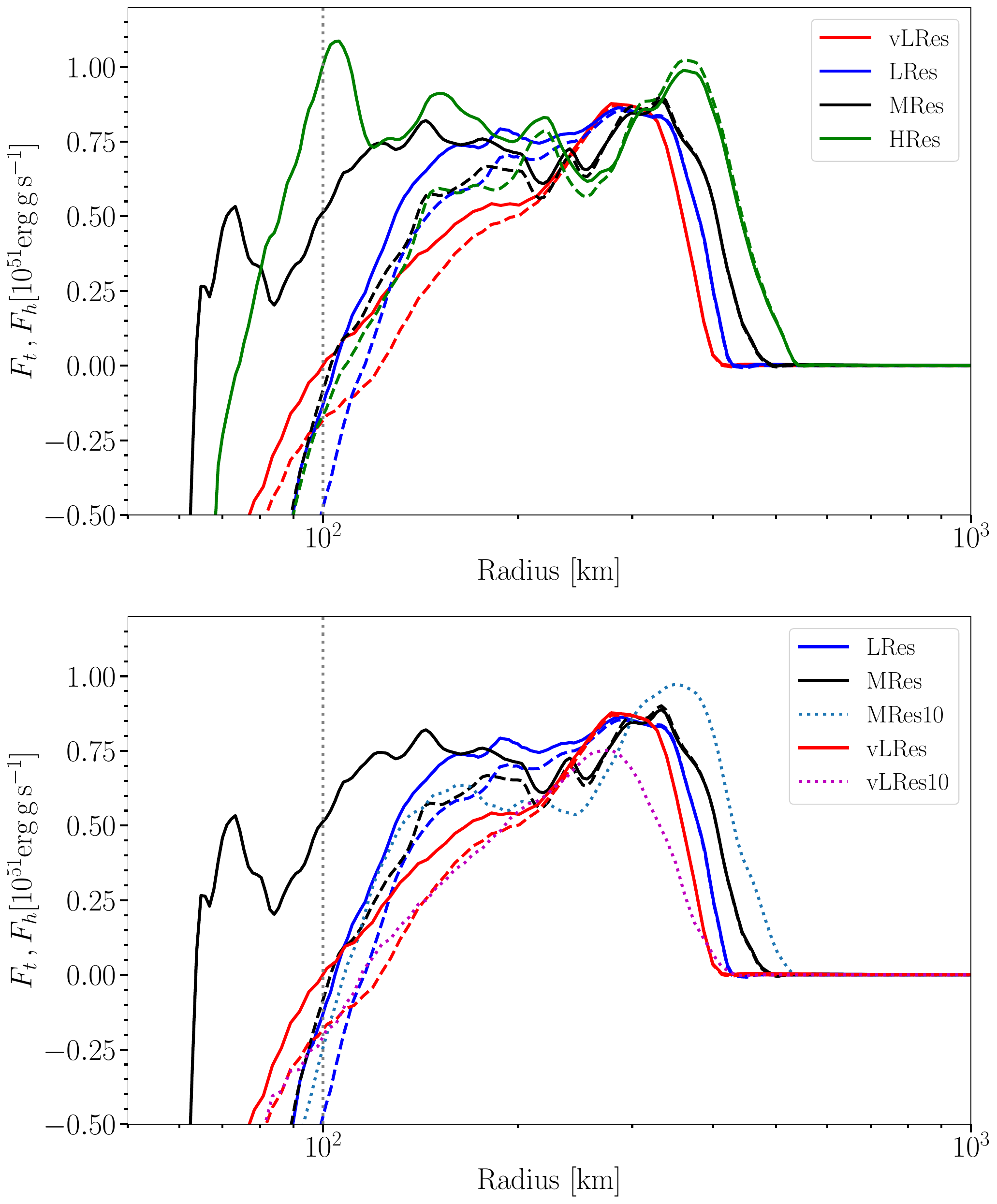}
 \caption{\normalsize Hydrodynamic enthalpy flux ($F_\mathrm{h}$, dashed) and total energy flux ($F_\mathrm{t}$, solid) at 0.20\,s post-bounce up to 1000\,km. The vertical dashed line gives the approximate radius ($\approx$100\,km) above which the angle-averaged total energy in the outflows becomes positive. }
 \label{fig:Flux}  
\end{figure}

As in \citet{Varma2023}, we try to understand the energy contributions to the diagnostic explosion energy by comparing fluxes that characterise energy transport in the gain region in Figure~\ref{fig:Flux}. To disentangle the role of magnetic fields, we plot the total magnetohydrodynamic flux, $F_\mathrm{t}$, which is defined as,
\begin{equation}
F_\mathrm{t} = \int\limits_{v_r>0}  r^2 
\left[(\rho e+P_\mathrm{t}+\rho \Phi) v_r + \frac{\mathbf{B} (\mathbf{v}\cdot\mathbf{B})}{4\pi} \right]
\,\ud \Omega ,
\end{equation}
and compare this to the purely hydrodynamic total enthalpy flux $F_\mathrm{h}$,
\begin{equation}
F_\mathrm{h} = \int\limits_{v_r>0} r^2 
(\rho e+P+\rho \Phi) v_r 
\,\ud \Omega,
\end{equation}
where the magnetic pressure and magnetic stresses are excluded and only
the gas pressure $P$ enters aside from purely advective terms. 

The fluxes $F_\mathrm{t}$ are plotted with solid lines, and the equivalent $F_\mathrm{h}$ with dashed lines. While magnetic fields are present for vLRes10 and MRes10, their magnetic contributions to the fluxes are negligible. We plot their total flux in dotted lines in the bottom plot of Figure~\ref{fig:Flux}.  

In Figure~\ref{fig:Flux} (top), we compare the four different resolution models (vLRes, LRes, MRes, HRes) with initial magnetic fields of $B = 10^{12}\,\mathrm{G}$.
As explained above, the higher resolution models explode slightly faster, so, at any snapshot in time, the positions of features of the fluxes (e.g., the shock) will differ slightly. We pick a time of 0.20\,s, where these models have all just entered runaway expansion, but are still very close in total explosion energy to make the comparison more meaningful. Given that, it is clear that the total flux contributions, $F_\mathrm{t}$, in the inner regions of the star (i.e. $\approx$100\,km - 200\,km) correlate with grid resolution, as we see an increase in $F_\mathrm{t}$ in this radial extent as we increase our grid resolution.

Unlike $F_\mathrm{t}$, the enthalpy fluxes, $F_\mathrm{h}$ of LRes, MRes and HRes are all very similar to each other at this time. The difference in $F_\mathrm{t}$ between these three models implies that the additional turbulent energy from more resolved grids is largely converted into magnetic energy below 200\,km. We note that likely due to the higher numerical diffusion, model vLRes does not follow this trend. 
The enthalpy flux for vLRes is slightly lower between $\approx$100\,km - 200\,km than the other models. There is still an increase in $F_\mathrm{t}$ compared to $F_\mathrm{h}$ in the inner region of this model due to the additional impact of the magnetic terms, however, its impact is less noticeable than the models with more resolved grids.

Although we see that the magnetic components of $F_\mathrm{t}$ can dominate the flux around the gain radius, even for our highest resolution mode, HRes, the purely hydrodynamic terms are dominant outside 250\,km. $F_\mathrm{h}$ is hence dominant at the shock interface for all our models, and ultimately drives the revived shock. Given the growing fluxes due to the magnetic terms in the gain region, it is unclear if the magnetic fluxes would eventually drive the shock if there was more time before shock revival was achieved.

When MRes and vLRes are compared to their weakly magnetic counterparts, MRes10 and vLRes10, in the bottom plot of Figure~\ref{fig:Flux}, the enthalpy fluxes between the differently magnetised models are qualitatively similar. This implies that the increase in explosion energy between these models is likely directly related to the magnetic contributions of the more strongly magnetised simulations. 

Since the explosion energies of LRes and MRes10 follow a very similar trajectory, we also compare their flux contributions. We note that LRes has a higher $F_\mathrm{t}$ in the inner ejecta regions, whereas the flux of MRes10 at the shock front is even higher than that of MRes. The higher $F_\mathrm{t}$ in MRes10 is due to a higher turbulent kinetic energy flux. Although MRes has the same grid resolution, and hence should have the same turbulent kinetic energy, some of this energy is converted to magnetic energy through a turbulent dynamo. The LRes model, on the other hand, is more numerically diffusive but has a stronger initial magnetic field, which leads to an integrated flux that is very similar to MRes10.

Due to the early explosion, our results for this 13$M_\odot$ progenitor appear to have a smaller dependence on the initial magnetic field strength than seen previously \citep{MullerVarma2020, Varma2023}. Here we find that this secondary effect of a strong magnetic field has a similar magnitude effect on the early diagnostic explosion energy as a change in the angular resolution of the simulation (e.g. MRes10 and LRes).

\subsection{Neutrino and Gravitational Wave Emission}
\label{subsec:Neutrino}

\begin{figure*}
\centering
    \subfloat{\includegraphics[width=0.5\linewidth]{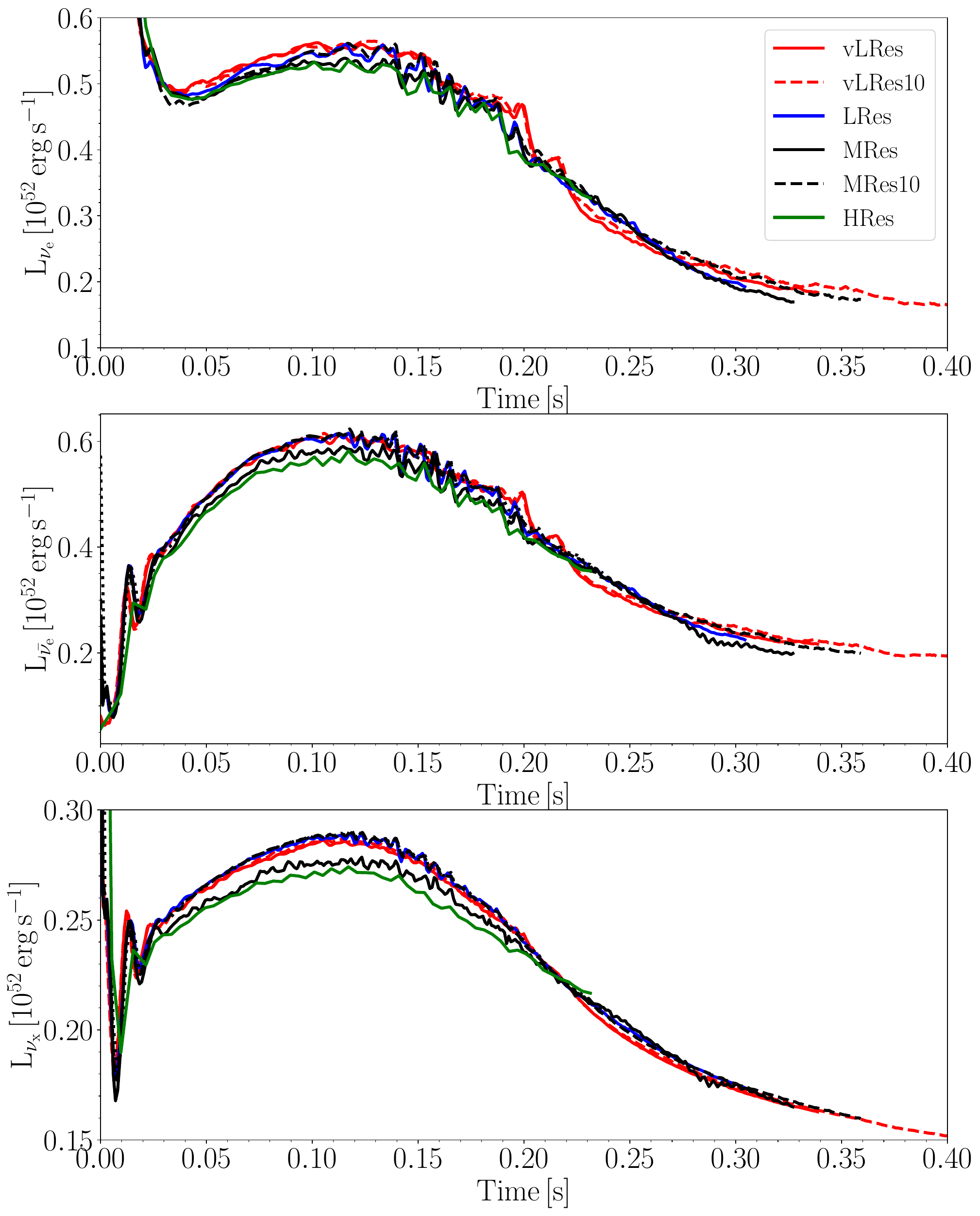}}
\hfil
    \subfloat{\includegraphics[width=0.5\linewidth]{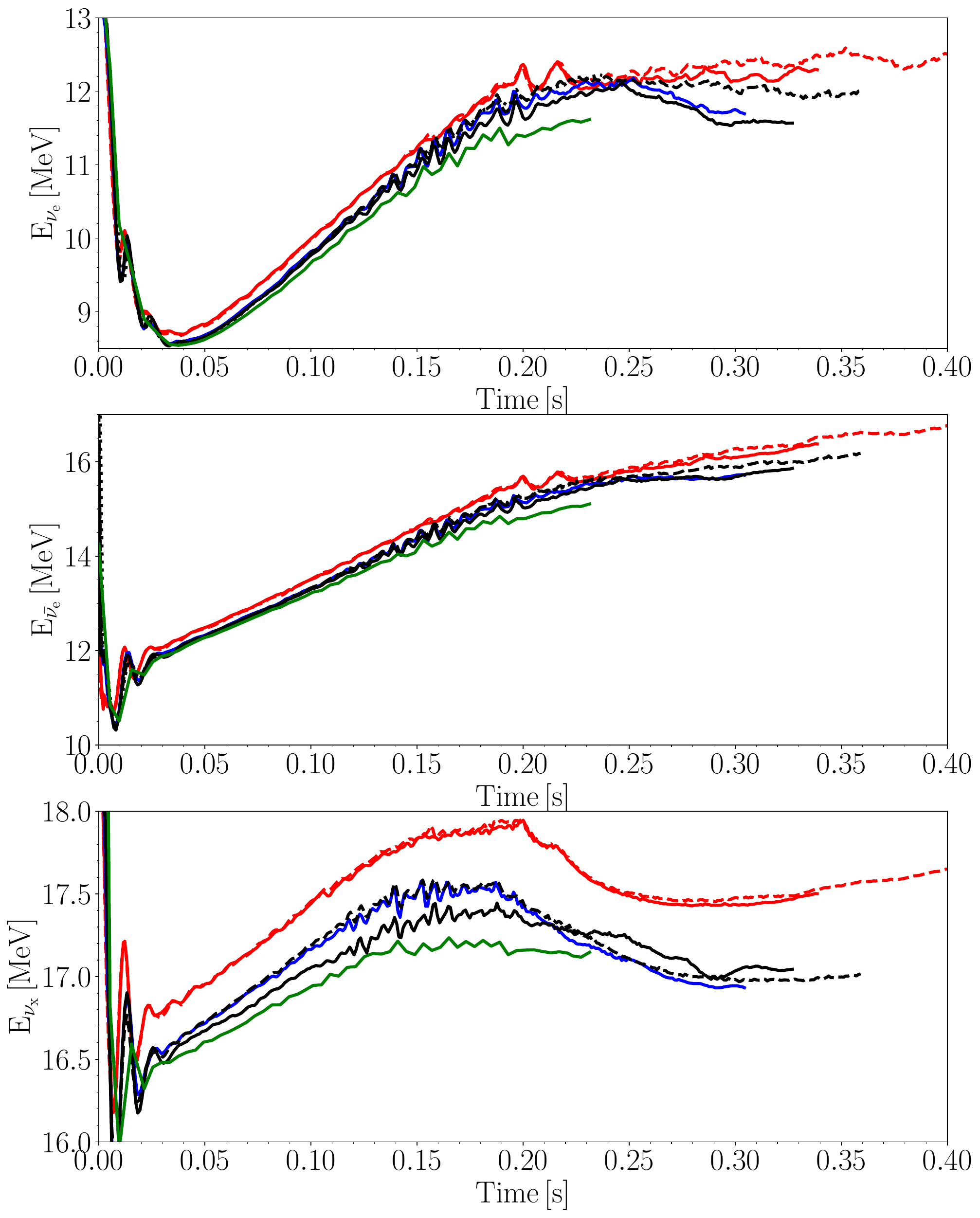}}
\caption{\normalsize Neutrino luminosity (left) and neutrino mean energy (right) for electron neutrinos, $\nu_\mathrm{e}$ (top), electron antineutrinos, $\bar{\nu}_\mathrm{e}$(middle), and heavy-flavour neutrinos, $\nu_x$(bottom), respectively.  }
    \label{fig:Luminosity}
\end{figure*}

As discussed in Section~\ref{subsec:dynamics}, in our strongly magnetic models, we find a correlation between the spatial resolution of our simulations and the bulk properties of the explosion. Since our explosions are still primarily neutrino-driven, we begin investigating this relation by analysing the neutrino light curves and mean energies in Figure~\ref{fig:Luminosity}.

Figure~\ref{fig:Luminosity} shows the evolution of the neutrino luminosities and energies for electron neutrinos, $\nu_\mathrm{e}$, electron antineutrinos, $\bar{\nu}_\mathrm{e}$, and heavy-flavour neutrinos, $\nu_x$. We find that in our models, higher resolution \emph{does} impact the neutrino emission. 
Unintuitively, given that there is a positive correlation between the resolution of our models and the explosion energy, we find an inverse relationship between resolution and neutrino luminosities and energies for all flavours, after around 0.03\,s post-bounce ($\mathord{\approx}8$\% between the lowest and highest resolutions). We see indirect evidence here that the lower neutrino luminosities and energies are caused by stronger magnetic fields. As the resolution of the model increases, the rate of magnetic field amplification in the fluid by a small-scale turbulent dynamo becomes more efficient, hence the higher-resolution models have systematically stronger magnetic fields. This is supported by comparing the strongly magnetic model, MRes, which has clearly lower neutrino luminosities and energies by $\mathord{\approx} 5$\% for all flavours compared to its weakly magnetised counterpart, MRes10. Prior to shock revival ($\approx$0.20\,s), this trend appears to hold, aside from the heavy-flavour neutrino luminosities of LRes, which are slightly higher than those seen in vLRes. Similar to the time evolution of the explosion energy (Figure~\ref{fig:Exp_shock}), both LRes and MRes10 show strikingly similar neutrino luminosities and energies. 

Previous resolution studies for hydrodynamic CCSNe \citep[e.g.,][]{Nagakura2019} do not exhibit any strong difference in the neutrino luminosity or energies at the different resolutions\footnote{Although a similar resolution study was performed by \citet{Melson2020}, due to their more simplified treatment of neutrino emission, using a light bulb model, they could not have observed these differences in neutrino luminosity that we note in our models.}, changes only occur in their models after revival due to changes in the accretion onto the PNS. 
The faster, more energetic explosions seen in pure hydrodynamics models at higher resolutions are instead attributed to an increase in turbulent pressure behind the shock since the energy provided by neutrino luminosity is unchanged. The trends seen in our models imply that the resolution effect of these neutrino properties is caused by a secondary effect of the presence of strong magnetic fields.

It is also of interest to note that the two very low-resolution models (vLRes and vLRes10) show very similar neutrino properties in Figure~\ref{fig:Luminosity} despite the difference in initial magnetic fields. This is likely due to the very high numerical diffusion at these low resolutions, making it difficult for these models to sustain strong magnetic fields. This indicates that if the spatial resolution is too low, the presence of magnetic fields may not be able to impact the explosion mechanism in a meaningful way.

\begin{figure}
 \includegraphics[width=\columnwidth]{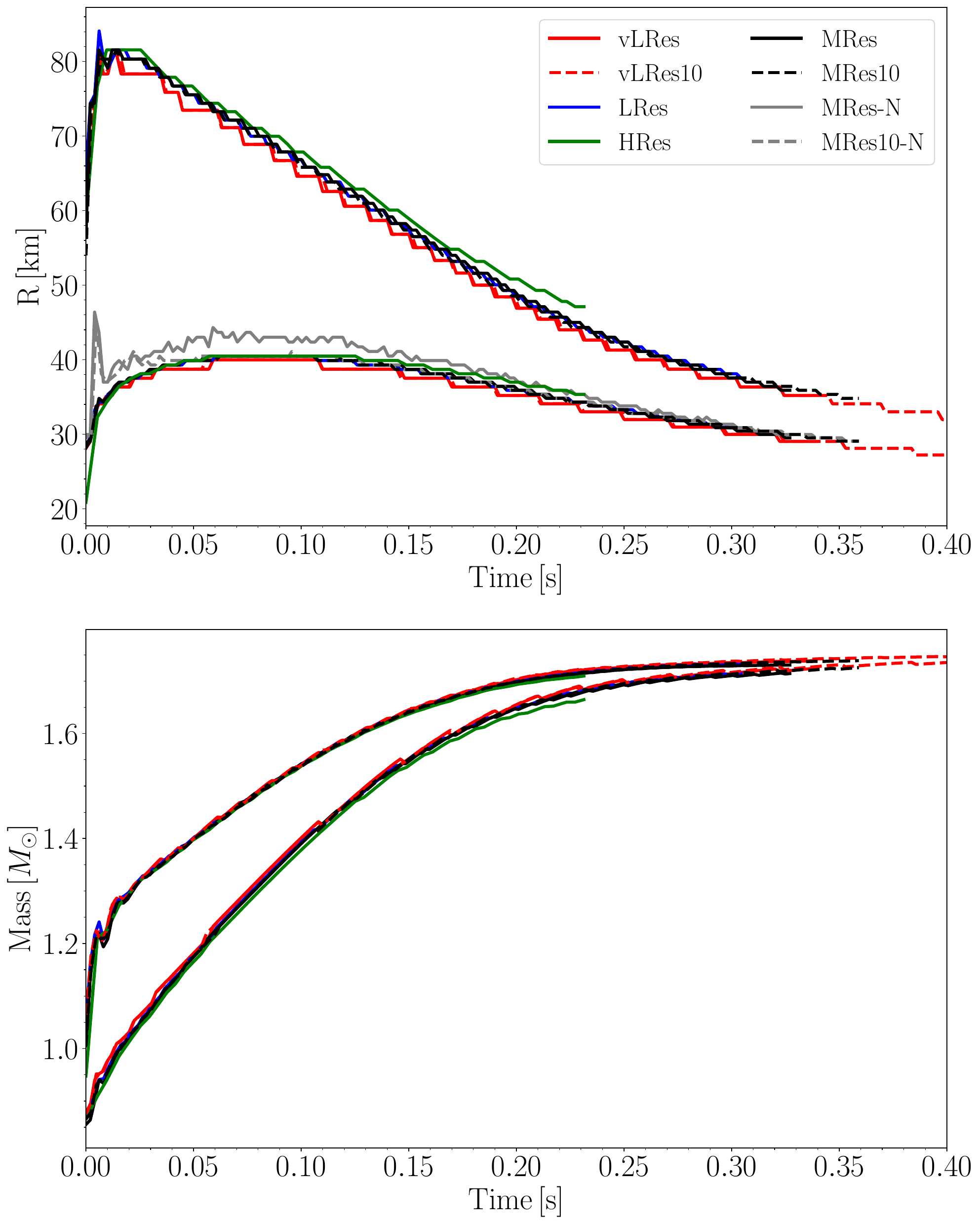}
 \caption{\normalsize Evolution of the spherically averaged PNS radius (top) and mass (bottom). The two sets of models are for two choices for the definition of the PNS surface, $\rho = 10^{11}\mathrm{g\,cm^{-3}}$ and $\rho = 10^{12}\mathrm{g\,cm^{-3}}$. The grey lines, MRes-N and MRes10-N, in the PNS radius plot represent the radius of MRes and MRes10 at the ''North'' pole ($\theta=0$) corresponding to  $\rho =10^{12}\mathrm{g\,cm^{-3}}$.}
 \label{fig:PNS}  
\end{figure}

We expect that the neutrino luminosities (and energies) are directly related to conditions in the forming PNS \citep{Mirizzi2015, Muller2019b}, such that,

\begin{equation}
    L_{\nu_\mathrm{e}} + L_{\bar{\nu}_\mathrm{e}} = 2\beta_1 L_{\nu_x} + \beta_2 \frac{GM_\mathrm{PNS}\dot{M}}{R_\mathrm{PNS}}, 
\end{equation}

and,

\begin{equation}
    L_{\nu_x}  = 4\pi \phi \sigma_\nu {R_\mathrm{PNS}}^2 {T_\nu}^4, 
\end{equation}
where $M_\mathrm{PNS}$ is the mass of the forming proto-neutron star, and $R_\mathrm{PNS}$ is its corresponding radius, $\phi$ is a "greyness factor", $T_\nu$ is the temperature at the neutrinosphere, and the constant $\sigma_\nu$ is the Stephan-Boltzmann constant for neutrinos.
To understand the differences in neutrino emission between these models, we must understand how changes in resolution, and the subsequent differences in magnetic field strength affect the PNS structure and evolution. Figure~\ref{fig:PNS} shows the evolution of the radius, mass and magnetic energy of the forming PNS. We consider two choices for where the PNS surface is defined. For the radius and mass of the PNS in Figure~\ref{fig:PNS}, we plot both the region where the density, $\rho>\mathrm{10^{11}g\,cm^{-3}}$ and where $\rho>\mathrm{10^{12}g\,cm^{-3}}$.

For both the PNS radius and mass, we find only a weak dependence on resolution at both density choices and see that after shock revival. There is an indication of a slightly
larger radius for $\rho=\mathrm{10^{11}g\,cm^{-3}}$ 
and smaller PNS mass in model HRres and smaller radius 
and larger PNS mass in model vLRes, which is consistent
with a lower neutrino luminosity in HRes.
However, the PNS masses and radii of models LRes, MRes and MRes10 all consistently lie on top of one another throughout the evolution. While this similarity may show us why the neutrino properties of LRes and MRes10 are so similar, it does not explain the clear differences in neutrino luminosity and energy we see between MRes and MRes10 in Figure~\ref{fig:Luminosity}. These small differences in PNS structure likely also interact with a genuine resolution dependence of the neutrino transport solver.

\begin{figure}
 \includegraphics[width=\columnwidth]{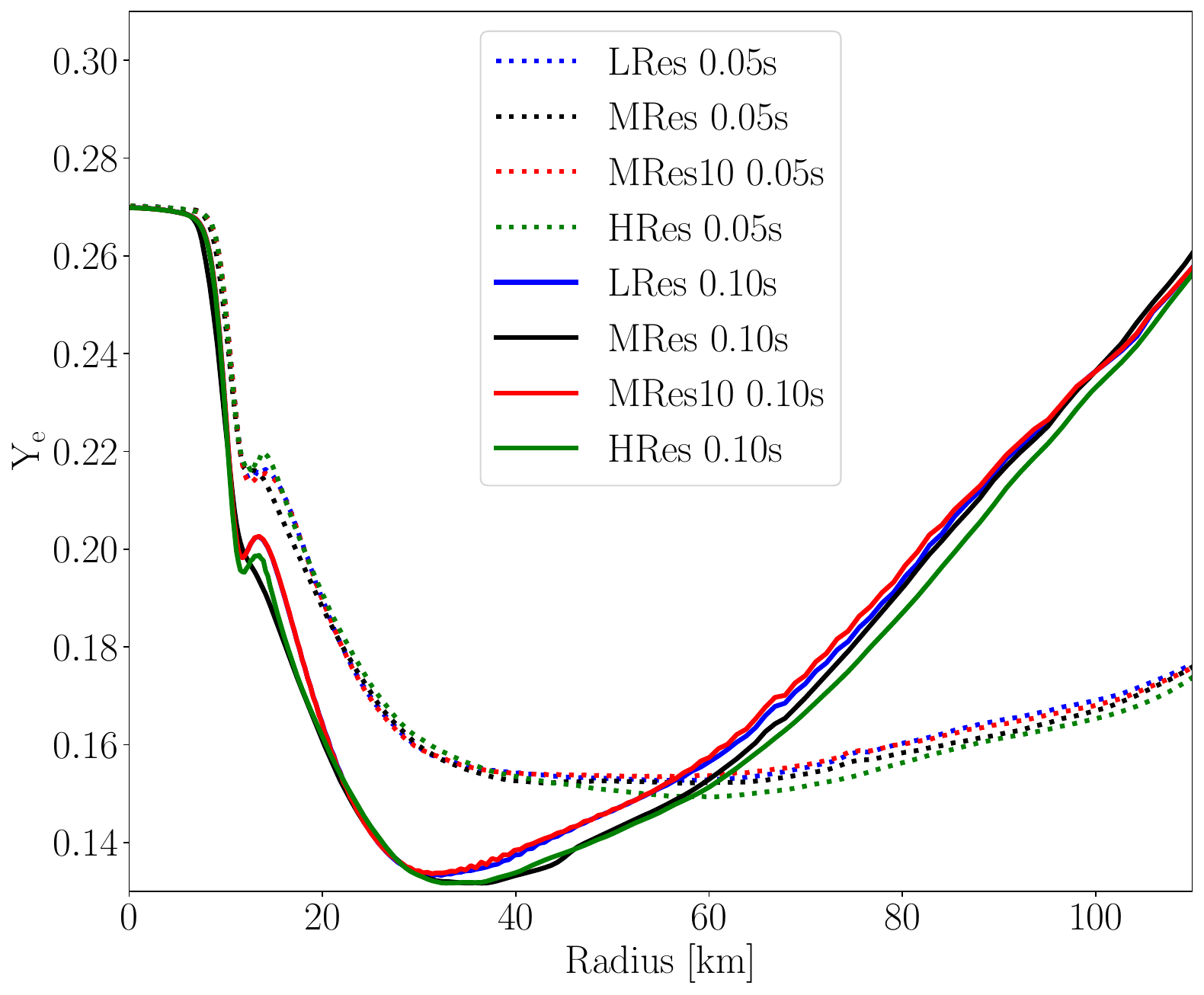}
 \caption{\normalsize Angle-averaged electron fraction $Y_\mathrm{e}$ in the core at 0.05\,s and 0.10\,s.}
 \label{fig:Ye}  
\end{figure}

\begin{figure}
 \includegraphics[width=\linewidth]{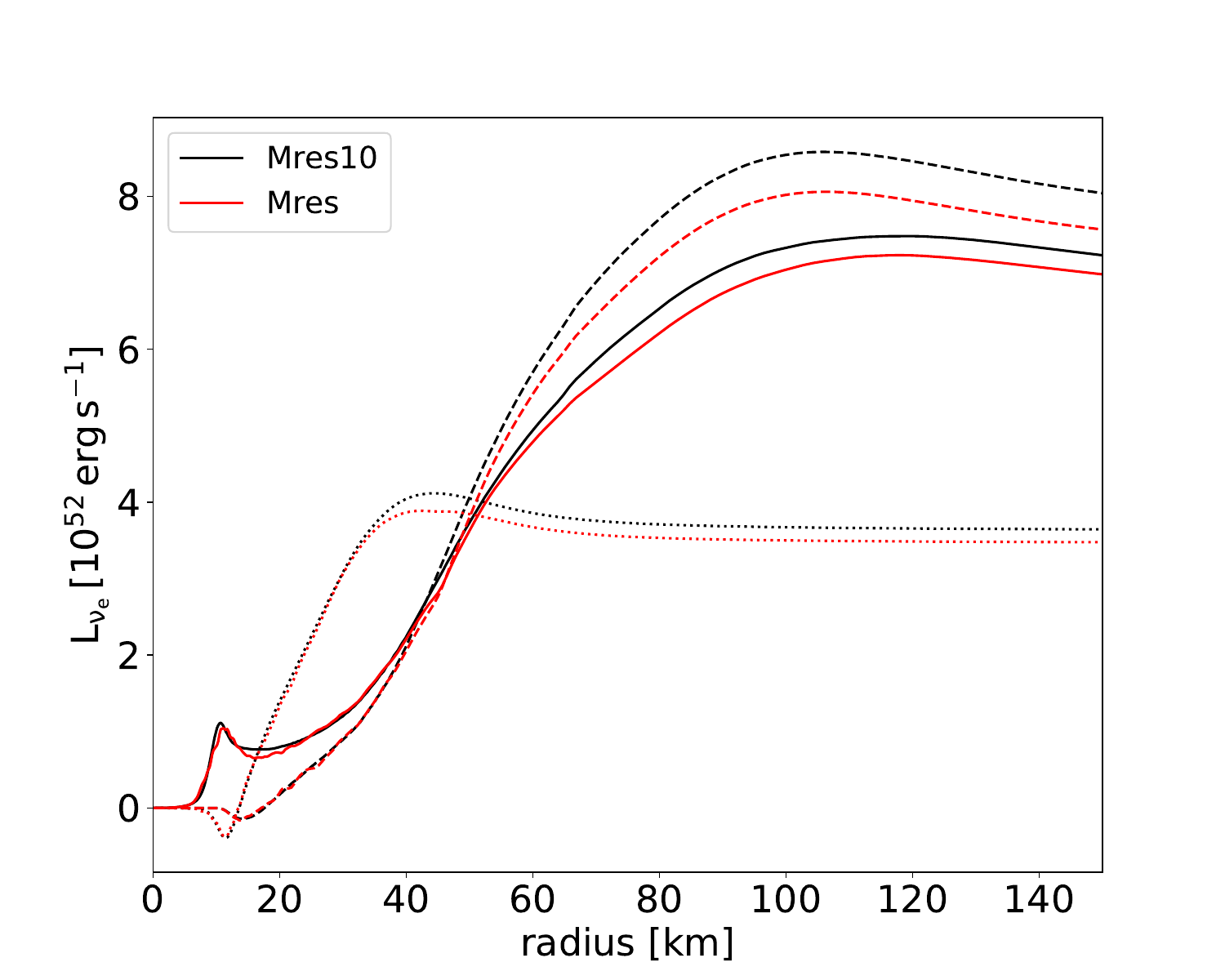}
 \caption{\normalsize Radial profiles of $L_{\nu_\mathrm{e}}$ (solid), $L_{\bar{\nu}_\mathrm{e}}$ (dashed) and $L_{\nu_x}$ (dotted), up to 150\,km for models MRes (red) and MRes10 (black) at 0.10\,s.
 }
 \label{fig:nu_rad}  
\end{figure}

The difference in neutrino emission between MRes and MRres10
can be partly understood by recognising that the strongly magnetised PNS (top plot of Figure~\ref{fig:PNS}) is distorted by the strong dipole magnetic field. To show this, we plot the radius of the PNS at the \emph{``North'' pole}, where $\rho>\mathrm{10^{12}g\,cm^{-3}}$ and $\Theta\,=0^\circ$, in grey in Figure~\ref{fig:PNS} for MRes and MRes10. Very quickly (i.e. $\approx0.04$\,s), we find that the radius at the pole for MRes10 falls and is then equivalent to its angle-averaged value. Although both models have almost indistinguishable angle-averaged $R_\mathrm{PNS}$, MRes has a radius at the pole that is up to 5\,km larger, leading to lower neutrino luminosities and energies, since some of the neutrinos are released from larger radii and hence lower temperature regions. As the PNS evolves, $R_\mathrm{PNS}$ for MRes becomes more spherically symmetric, with the radius at the pole eventually becoming equivalent to that in MRes10 by $\approx$0.25\,s. Figure~\ref{fig:Luminosity} shows how this, in general, leads to the neutrino luminosities of MRes and MRes10 becoming more similar. 

We find that on top of this, the different neutrino emission in the cooling region impacts the $Y_\mathrm{e}$ distribution. Figure~\ref{fig:Ye} shows $Y_\mathrm{e}$ profiles at 0.05\,s and 0.10\,s for LRes, MRes, MRes10 and HRes. While all four models are largely similar at 0.05\,s (with slightly lower values $\mathord{>}$50\,km for HRes), their distributions have visually changed by 0.10\,s as the core continues to contract. While LRes and MRes10 appear to have a largely similar distribution, the higher-resolution and more strongly magnetised models show a drop in $Y_\mathrm{e}$ between 35\,km and 80\,km for MRes and up to 100\,km for HRes. 
In Figure~\ref{fig:nu_rad}, we show the radial profiles of $L_{\nu_\mathrm{e}}$, $L_{\bar{\nu}_\mathrm{e}}$ and $L_{\nu_x}$ for models MRes and MRes10 at 0.10\,s post-bounce. We see that the two models start deviating at $\approx$40\,km due to the inflated, and slightly lower temperature of the PNS in MRes. 
This is a strong indication that the different neutrino emission is due to different emission rates in the region between 40\,km and
80\,km, which is then also bound to shift the quasi-equilibrium between neutrino emission and absorption that determines the $Y_\mathrm{e}$ in this region.
\begin{figure*}
\centering
    \begin{subfigure}{0.49\linewidth}
        \includegraphics[width=\linewidth]{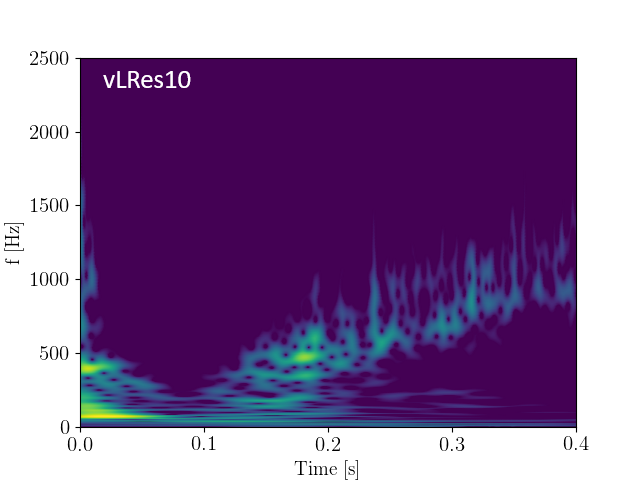}
    \end{subfigure}
\hfil
    \begin{subfigure}{0.49\linewidth}
        \includegraphics[width=\linewidth]{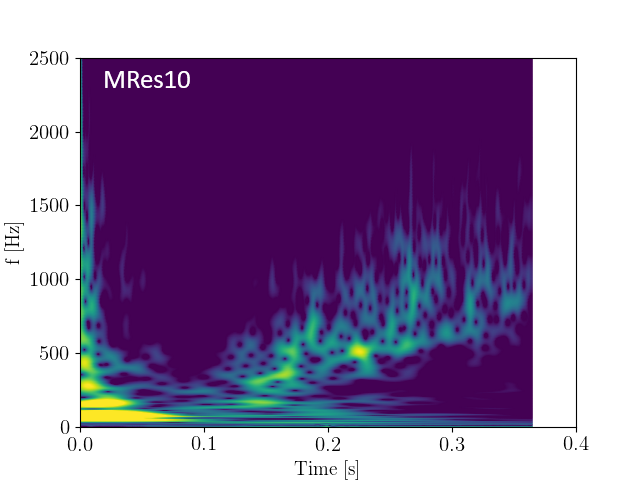}
    \end{subfigure}

    \begin{subfigure}{0.49\linewidth}
        \includegraphics[width=\linewidth]{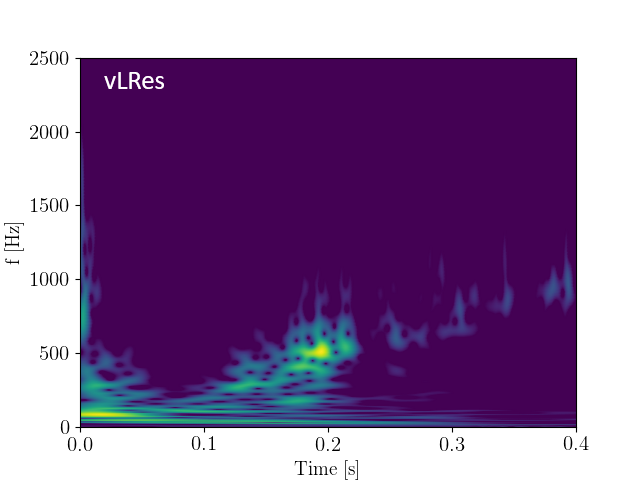}
    \end{subfigure}
\hfil
    \begin{subfigure}{0.49\linewidth}
        \includegraphics[width=\linewidth]{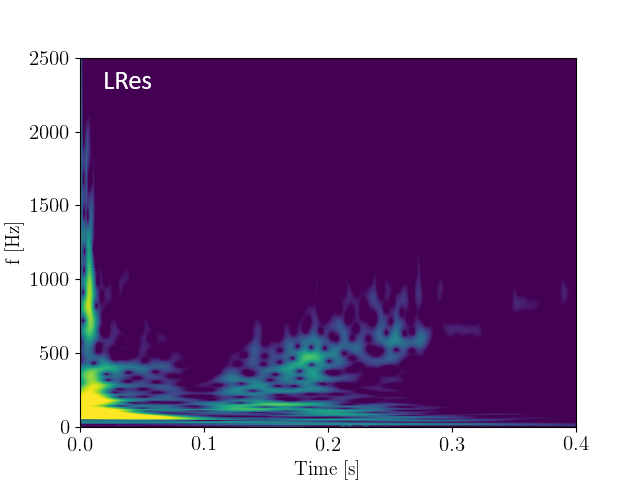}
    \end{subfigure}

    \begin{subfigure}{0.49\linewidth}
        \includegraphics[width=\linewidth]{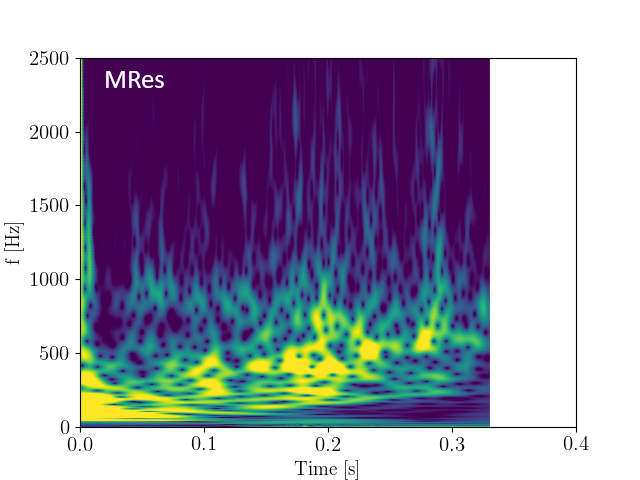}
    \end{subfigure}
\hfil
    \begin{subfigure}{0.49\linewidth}
        \includegraphics[width=\linewidth]{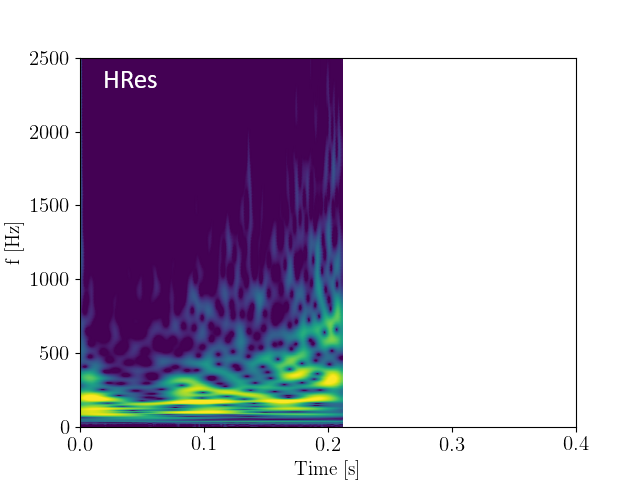}
    \end{subfigure}
\caption{Spectrograms of $h_+$ for each of our simulated models. The colour scale is consistent between each model.}
    \label{fig:GW}
\end{figure*}

Another important multi-messenger signal to analyse from this set of models is the gravitational wave (GW) signal. We present the spectrograms for each of our models in Figure~\ref{fig:GW}. Similar to the non-rotating models presented in \citet{Powell2024}, we first see a high-power hotspot at low frequency. This is the familiar signal from the ringdown from proto-neutron star convecction \citep{marek_09,yakunin_10,mueller_13}.

For our weakly magnetised models, vLRes10 and MRes10, after the ringdown, we start to see the usual ramp-up signal with a
rise from low to high frequencies starting from around 0.1\,s. This is also mirrored in our lower resolution models, with strong initial magnetic fields, vLRes and LRes. 

The inclusion of strong magnetic fields, when resolution is sufficient, sometimes changes the spectrogram qualitatively. Comparing MRes to MRes10, we see 
more power in the spectrogram as well as emission over a broader range of frequencies. The higher power is also seen in HRes. However, the situation is reversed in the comparison of  vLRes to vLRes10, which has less power in gravitational waves. Both these low-resolution models show less gravitational wave emission, consistent with higher numerical dissipation.

As additional feature in one of the strong-field models, we find a weaker emission band 
with decreasing frequency in model MRes, starting at around 1000Hz in MRes.
This corresponds to the core \emph{g}-mode signal identified in 2D by \citet{jakobus_23}. The core \emph{g}-mode signal was recently also seen in a magnetised 3D model, i.e., the 3D version of model SR1 in \citet{Sykes2025}. It is noteworthy that this signal has so far shown up only in magnetised models in 3D, so magnetohydrodynamic effects could have an impact on the excitation of this mode.
We speculate that the sufficiently strong magnetic fields in MRes could lead to a more efficient coupling between the PNS convection and the stable zone. However, the core $g$-mode signal does not appear in all of the MHD models, so the reasons for its appearance remain unclear and need to be investigated by further 3D studies.

\subsection{Gain Region Dynamics}
\label{subsec:Gain} 

\begin{figure}
 \includegraphics[width=\columnwidth]{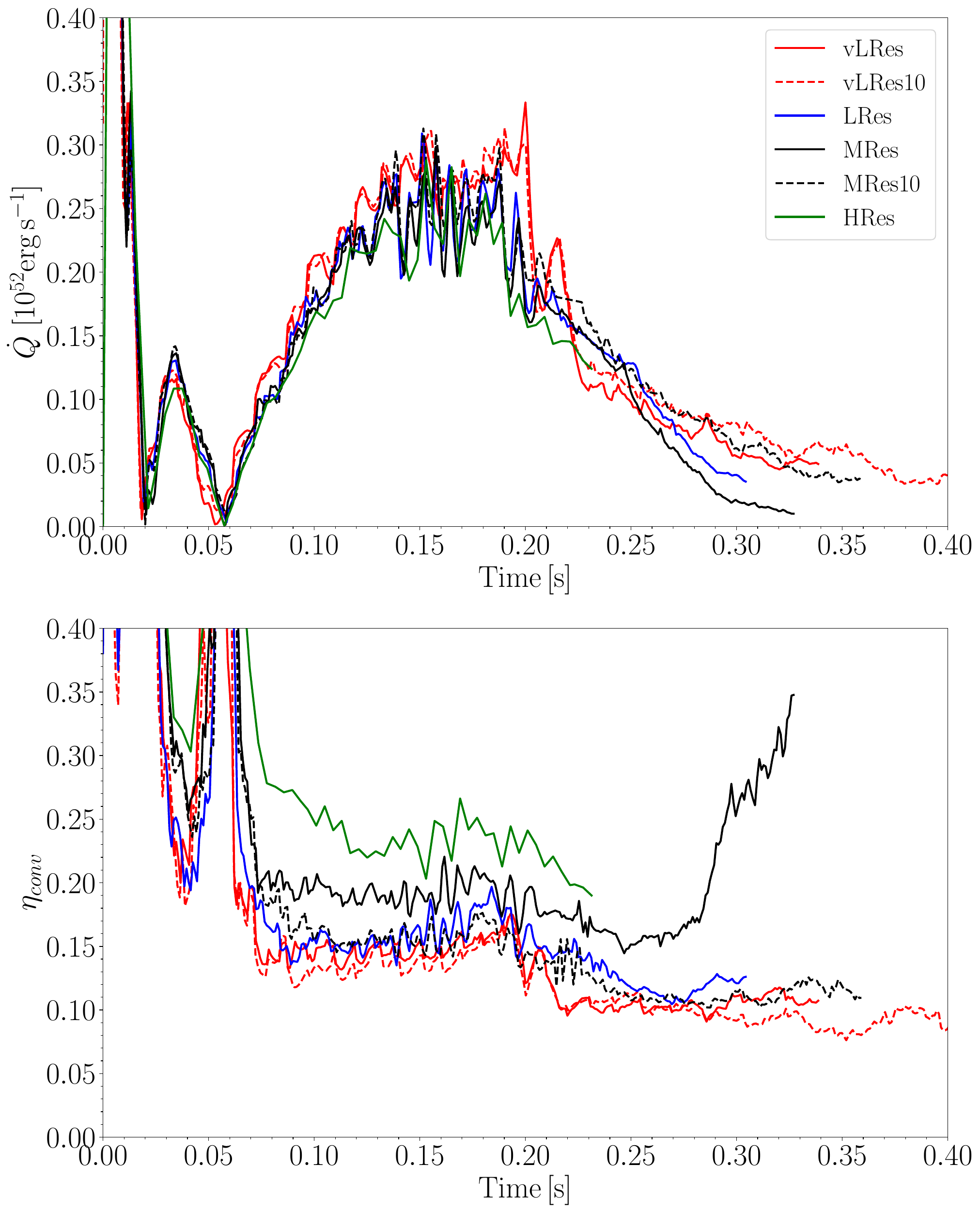}
 \caption{\normalsize Time evolution of neutrino heating rate, $\dot{Q}_\nu$ (top), and the convective efficiency, $\eta_\mathrm{conv}$ (bottom).}
 \label{fig:Nheat}  
\end{figure}

To understand how the magnetic fields aid the neutrino-driven mechanism at higher resolutions and how the strong magnetic contributions to the fluxes develop (Figure~\ref{fig:Flux}), we turn our attention to the dynamics in the gain region.  
Their impact inside the gain region can be probed, in part, by the neutrino-heating rate, $\dot{Q}_\nu$. The volume-integrated heating rate in Figure~\ref{fig:Nheat} (top) quantifies the neutrino heating rate in the gain region. Here, we find an inverse relationship between resolution and $\dot{Q}_\nu$. This is unsurprising, given the inverse correlation between resolution and both neutrino luminosity and mean energies that we noted in Section~\ref{subsec:Neutrino}.

The heating rate for models MRes and MRes10 is slightly difficult to distinguish visually before runaway shock expansion is achieved at $\approx$0.20\,s, due to the stochastic fluctuations. However, between 0.13\,s and 0.20\,s, the heating rate for MRes is slightly lower than MRes10, with slightly lower dips and slightly higher peaks in the heating rates for MRes10. Similarly, for the very low-resolution models, vLRes has slightly lower heating than vLRes10 between 0.15\,s and 0.20\,s. The heating rate of LRes is even more difficult to distinguish from MRes10 prior to 0.20\,s. 

Part of the discrepancy can be resolved by noting the convective efficiency parameter, $\eta_\mathrm{conv}$, which quantifies the efficiency of the conversion of neutrino heating to turbulent kinetic and magnetic energy in the gain region, \citep{Mueller2015,MullerVarma2020}, which is defined as, 

\begin{equation}
    \eta_\mathrm{conv} = \frac{(E_\mathrm{kin} + E_\mathrm{{B }})/M_\mathrm{gain}}{[(r_\mathrm{sh}-r_\mathrm{gain})\dot{Q}_{\nu}/M_\mathrm{gain}]^{2/3}}.
    \label{eq:Nconv}
\end{equation}

Despite the lower heating rates (Figure~\ref{eq:Nconv}), there is a clear trend, with up to 50\% higher efficiency at higher resolution. The higher resolution models are more efficient at converting the available heating energy into turbulent kinetic energy. The $ \mathrm{30\%}$ lower $\eta_\mathrm{conv}$ seen in MRes10 compared to MRes indicates that a stronger magnetic field is also contributing to the increased efficiency. Once again, the very high numerical diffusivity of the vLRes and vLRes10 lead to their $\eta_\mathrm{conv}$ parameters being largely similar. There is, however, a slightly higher efficiency in vLRes between 0.09\,s and 0.18\,s, likely due to the initially higher magnetic field strength.  

\begin{figure}
 \includegraphics[width=\columnwidth]{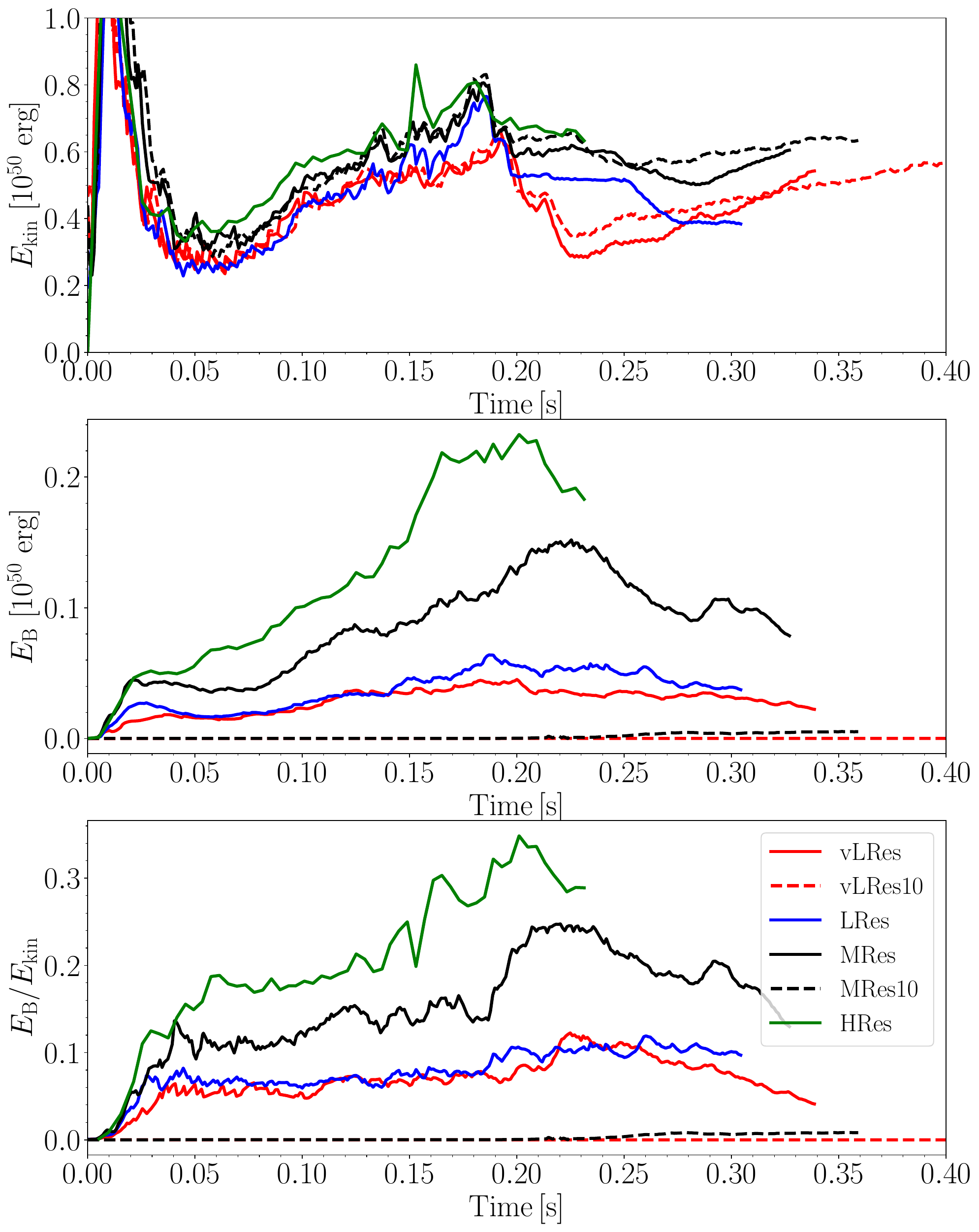}
 \caption{\normalsize Time evolution of kinetic energy, $E_\mathrm{kin}$ (top), magnetic energy, $E_{B}$ (middle), and the ratio of kinetic to magnetic energy, $E_\mathrm{kin}/E_{B}$, in the gain region (bottom). }
 \label{fig:Gain}  
\end{figure}

To better understand $\eta_\mathrm{conv}$ in the gain region, we compare the turbulent kinetic energy, $E_\mathrm{kin}$, magnetic energy, $E_{B}$, and their ratios in Figure~\ref{fig:Gain}.
$E_\mathrm{kin}$ and $E_{B}$ are computed as

\begin{align}
E_\mathrm{kin}&=\int \frac{1}{2}\rho |\mathbf{v}'|^2 \, \ud V,
\label{eq:gainenergy1}
\\
E_B&=\int \frac{|\mathbf{B}|^2}{8\pi}  \,\ud V,
\label{eq:gainenergy2}
\end{align}
where $\mathbf{v'}$ denotes the fluctuations of the velocity
around its spherical Favre average (i.e. $\mathbf{v'} = \mathbf{v} - \langle \mathbf{v} \rangle$).

Given that our simulations are implicit large-eddy simulations (ILES), and viscosity and magnetic diffusivity are not included explicitly in our numerical scheme, we expect that the Reynolds number, Re, increases with resolution as $\mathrm{Re} \propto N_{r}^{4/3}$ \citep{Cristini2017}, leading to stronger turbulent flow at higher resolutions. This trend towards higher turbulent kinetic energy is seen clearly in Figure~\ref{fig:Gain} (top). At their peak values, $E_\mathrm{kin}$ of vLRes is $\approx\mathrm{25\%}$ lower than HRes.  
Due to their use of the same angular resolutions, it is not surprising that models LRes and vLRes show largely similar values of $E_\mathrm{kin}$, but start to differ more noticeably after 0.15\,s. The higher diffusivity of both vLRes and vLRes10 leads to a slower rise in  $E_\mathrm{kin}$ (Figure~\ref{fig:Gain}), and also a lower peak value. This slower growth in  $E_\mathrm{kin}$ (and $E_{B}$) for the very low-resolution models after 0.15\,s is what leads to a slightly later shock revival than we see in their higher resolution counterparts.

Since no viscosity, $\nu$, or magnetic resistivity, $\eta$, is explicitly added, we can assume that all of our models have an effective magnetic Prandtl number ($\mathrm{Pm}=\nu/\eta$), $\mathrm{Pm} \sim 1$ in the ILES approach \citep{federrath_11}. Studies of turbulent dynamos \citep{Schekochihin2007, Leidi2023} indicate that, for a fixed Pm, the growth rate of the dynamo increases with grid resolution. Unlike the turbulent dynamo in stellar convective shells, the supernova gain region introduces additional non-stationarity because of the accretion flow and the time-varying size of the convective region. However, the general trend should still apply.
At $0.01\,\mathrm{s}$ post-bounce, the difference in growth rates between LRes and MRes is already evident. We see a trend of higher magnetic energy growth with resolution, with the HRes model achieving the highest $E_{B}$ values before shock expansion. 

As the resolution increases, $E_{B}$, increases, with vLRes having $\approx\mathrm{85\%}$ lower magnetic energy than HRes. 
Unsurprisingly, the weakly magnetised models (vLRes10, MRes10) have values of $E_B$ that are initially much lower than those of the more highly magnetised models and are negligible for the dynamics of their explosion mechanism. The trajectories of $E_{B}$ for the weakly magnetised models can't be distinguished in Figure~\ref{fig:Gain}, as they are several orders of magnitude lower than those in the strongly magnetised models. However, we note that by 0.25\,s, MRes10 has a magnetic energy in the gain region that is already of the order $10^{48}$erg. This very rapid growth of the field implies that for a progenitor with a more delayed explosion, there may be enough time for even the weakly magnetised models to be amplified to be dynamically relevant. Due to its very low resolution, vLRes10 only grows its field to $10^{45}$erg, where it appears to stop amplifying. 

Aside from $E_{B}$ and $E_\mathrm{kin}$ increasing with resolution, the ratio of magnetic to kinetic energy, $E_{B}/E_\mathrm{kin}$, also increases, implying that the additional turbulent kinetic energy is preferentially converted into $E_{B}$ than $E_\mathrm{kin}$ at higher resolutions. 
In \citet{MullerVarma2020} and \citet{Varma2023}, magnetic supernovae achieved ratios $E_{B}/E_\mathrm{kin}$ of $40\%$ - $50\%$, which triggered the runaway shock expansion. Here, we find that $E_\mathrm{B}/E_\mathrm{kin}$ is $\approx12\%,\, 11\%,\, 25\%\, and \,35\%$ of equipartition respectively for the vLRes, LRes, MRes and HRes models. 

It is prudent here to compare the MRes model to the magnetised model of \citet{MullerVarma2020}, which was simulated with the same grid resolution. In MRes, the ratio of $E_B/E_\mathrm{kin}$ only achieves a peak of $25\%$, while \citet{MullerVarma2020} show a ratio of $50\%$ when runaway explosion is achieved, although they start with much weaker magnetic field strengths. The primary cause of this is the difference in when the runaway explosion occurs. In our 13$\mathrm{M_\odot}$ progenitor, the explosion develops early, regardless of the initial magnetic field strength. Here, $E_\mathrm{kin}$ is converted into $E_{B}$, but also does work to drive the shock. While in \citet{MullerVarma2020}, the progenitor is more difficult to explode, so $E_\mathrm{kin}$ is preferentially converted to $E_B$.

As mentioned above, we find that this 13\,$\mathrm{M_\odot}$ progenitor explodes easily, as seen from the similarities between models of different resolutions and initial magnetic field strengths. The increased magnetic energies in the gain region do not trigger the explosion, but aid in the post-revival dynamics of the explosion. At higher numerical diffusivities in vLRes and LRes, the energy ratios are held roughly constant between 0.05 and 0.07 before 0.20\,s. 

\begin{figure*}
\centering
    \subfloat{\includegraphics[width=0.5\linewidth]{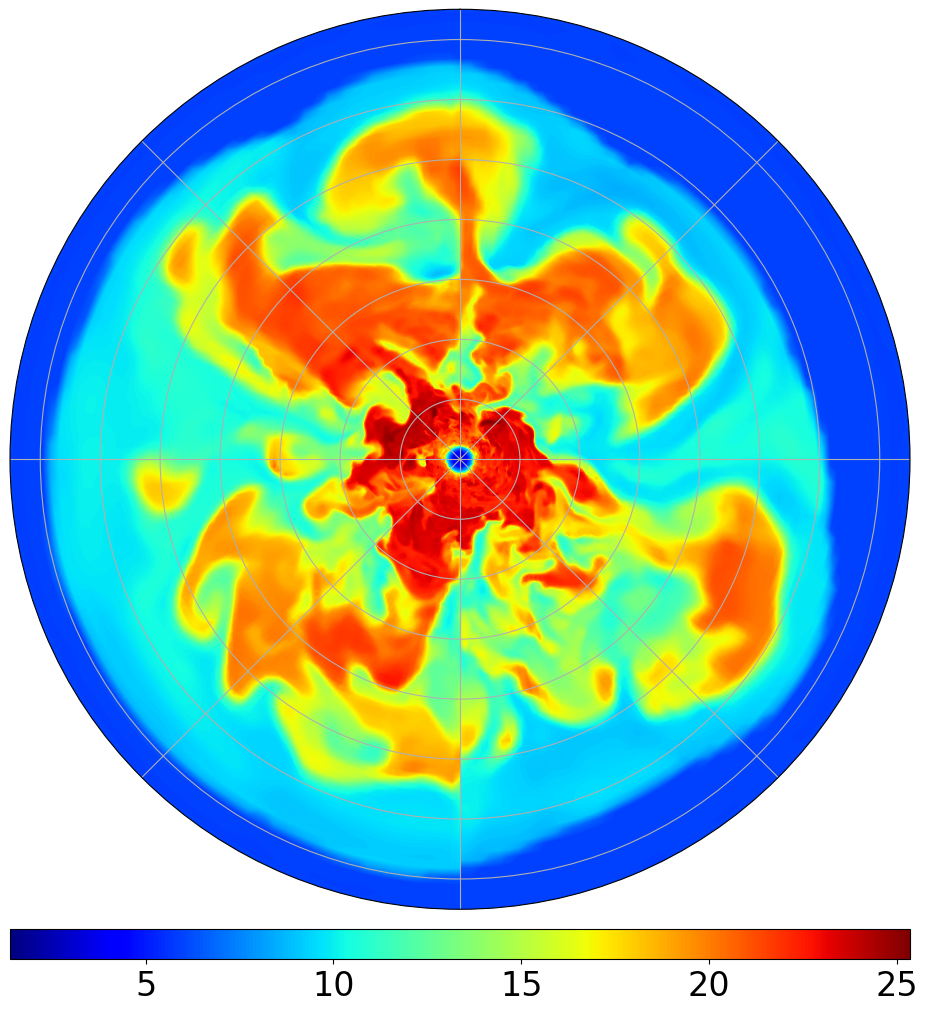}}
\hfill
    \subfloat{\includegraphics[width=0.5\linewidth]{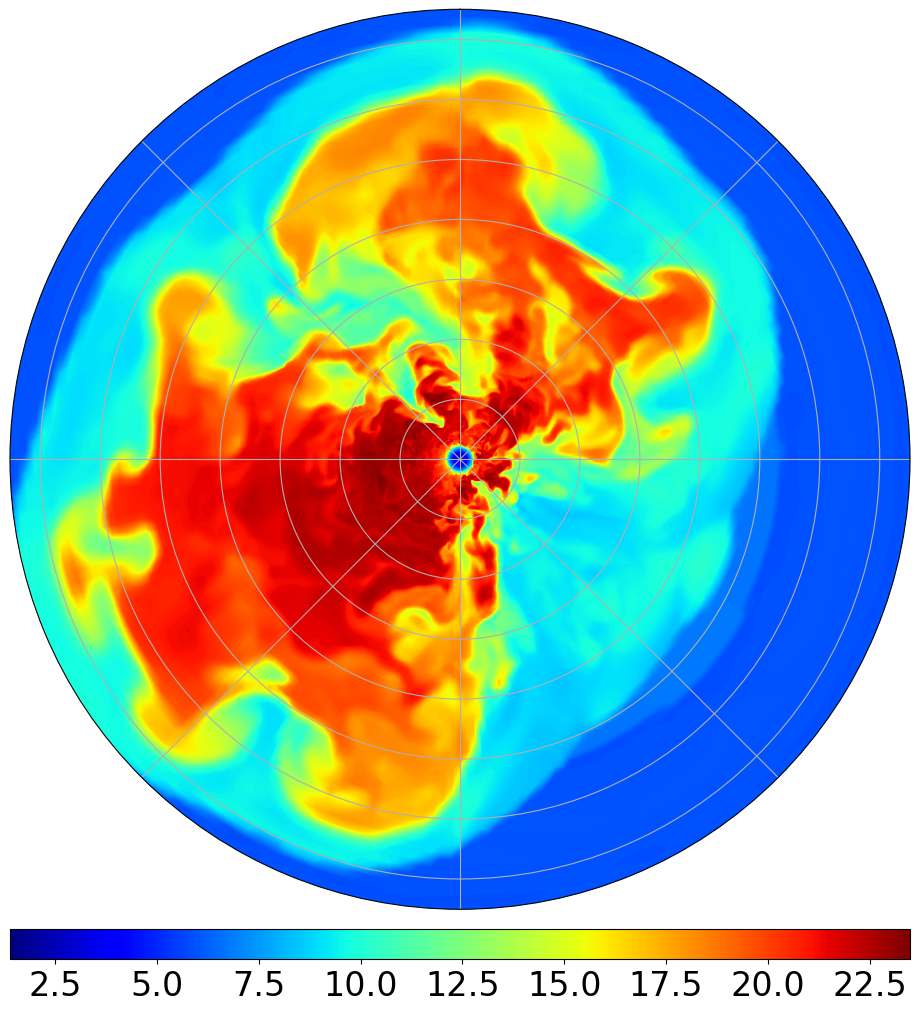}}
\caption{\normalsize Entropy on meridional slices through MRes and MRes10 models at 0.3\,s post-bounce out to 1500\,km. Each radial band in grey marks an additional 200\,km from the centre. }
    \label{fig:2D_morph}
\end{figure*}

As discussed in Section~\ref{subsec:dynamics}, although the increased magnetic energies do not directly drive the explosion, after shock revival, the models with higher $E_B$ have a larger total energy flux behind the shock, aiding the overall explosion dynamics. 
After shock revival, the rapid shock expansion leads to an initial drop in both $E_\mathrm{kin}$ and $E_B$. On the time scale of our simulations, the magnetic energies continue to drop in the gain region, however, after $\approx0.27$\,s, $E_\mathrm{kin}$ for most of these models either flattens or starts increasing once again. The relationships between the models of different resolutions and initial field strengths also appear to change. For example, at $\approx0.30$\,s, the vLRes model has increased its turbulent kinetic energy to now be above that of LRes. After 0.20\,s, the models with different initial magnetisations (vLRes and vLRes10, MRes and MRes10) now evolve independently of each other. This is due to differences in the morphology of the explosion (Figure~\ref{fig:2D_morph}), and changes in the PNS properties and evolution, which we will discuss in Section~\ref{subsec:PNS}. 
Importantly, the continued difference in $E_\mathrm{kin}$ and $E_B$ shows that our magnetic models have yet to converge, despite having qualitatively similar bulk explosion dynamics.

\subsection{Neutron star properties}
\label{subsec:PNS} 
We next consider the birth properties of the neutron stars produced in our suite of simulations and how they are impacted by the grid resolution. These issues are particularly relevant for evaluating the fossil-field scenario \citep{Spruit2009,ferrario_06, Makarenko2021} for magnetar formations. We stress that although we discuss the neutron star birth properties here, the magnetic field and spin rate can still be affected by processes that occur on longer time scales than we can simulate here, such as spin-up and field burial due to fallback \citep{Geppert1999, Ho2011, Vigano2012}.

\begin{figure*}
\centering
    \subfloat{\includegraphics[width=0.5\linewidth]{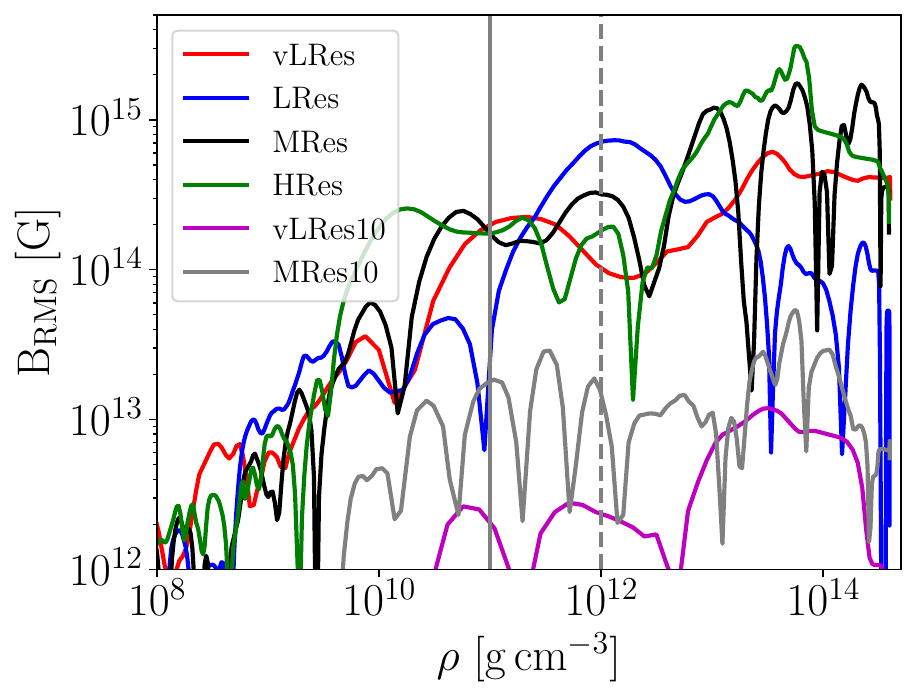}}
\hfil
    \subfloat{\includegraphics[width=0.5\linewidth]{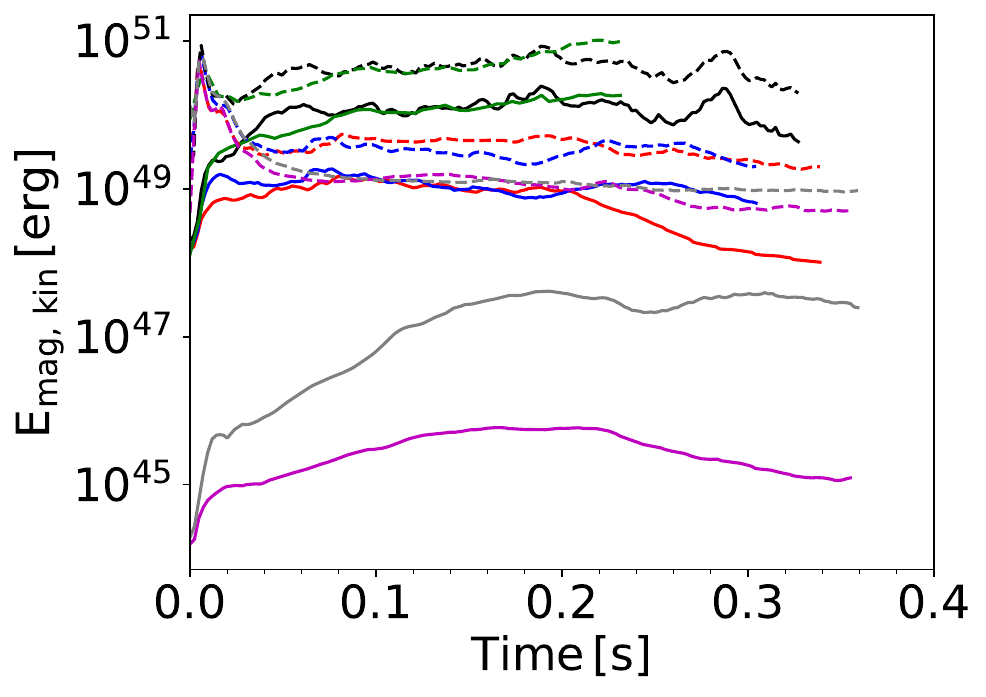}}
    \caption{\normalsize RMS magnetic field strength (see Equation~\ref{eq:Brms}) for all models as a function of density (left), and the magnetic and turbulent kinetic energy inside the PNS as a function of time (right). The PNS is defined as the region where $\rho>10^{11}\mathrm{g\,cm^{-3}}$. Grey lines in the left figure indicate densities of $\rho=10^{11}\mathrm{g\,cm^{-3}}$ and $\rho=10^{12}\mathrm{g\,cm^{-3}}$ (as often used to define the PNS  surface).}
    \label{fig:Mag}
\end{figure*}

We show the magnetic properties of the forming PNS in all our models in Figure~\ref{fig:Mag}, namely as profiles of the magnetic field strength against density at 0.23\,s, and the time evolution of the magnetic energy, $E_\mathrm{{mag}}$ inside the PNS. Where the RMS field strength is:
\begin{align}
\mathrm{B}_\mathrm{RMS}&=\int \frac{\mathbf{B}^2}{4\pi}  \,\ud \Omega.
\label{eq:Brms}
\end{align}
We define the PNS as the region where the density $\rho>\mathrm{10^{11}\,g\,cm^{-3}}$, but note that the time evolution of $E_\mathrm{{mag}}$ is not very sensitive to our choice for the PNS surface. Grey vertical lines are shown on our magnetic field strength profiles in Figure~\ref{fig:Mag} (left), to indicate the two common choices for PNS surface, $\rho=\mathrm{10^{11}g\,cm^{-3}}$ and $\rho=\mathrm{10^{12}g\,cm^{-3}}$. 

Magnetic field strength profiles show roughly similar PNS surface strengths at $\rho=\mathrm{10^{11}g\,cm^{-3}}$, with vLRes, MRes, and HRes all having values at $\approx2\times10^{14}$G. LRes, however, shows a sharp drop in its RMS field strength just at this boundary, dropping to $\approx3\times10^{13}$G. Just by choosing a different definition of the PNS surface, e.g. $\rho=\mathrm{10^{12}g\,cm^{-3}}$, we find that LRes now has the highest surface value, $\approx6\times10^{14}$G. One should not read too much into these individual values, as the PNS is still evolving, and the magnetic fields are being distributed by changes in the angular momentum redistribution and by PNS convection. Unlike in the models presented by \citet{Varma2023}, the PNS surface field strength here \emph{is} correlated with the initial field strength, which is also seen by the models presented by \citet{Matsumoto2022}. This difference could be due to the difference in initial magnetic field geometry, as \citet{Varma2023} uses a twisted torus initial field geometry, while, similar to \citet{Matsumoto2022}, we have used a simpler dipole configuration. A more detailed study of the impact of magnetic field geometries must be conducted to make any robust conclusions. 

Figure~\ref{fig:Mag} (right) shows the evolution of the integrated magnetic energy (solid lines) inside the forming neutron star, along with the turbulent kinetic energy, $E_{kin}$, (dashed lines) in this region. All our models show an initial period of very rapid magnetic field growth in the first 0.01\,s, largely due to the continued growth of $M_\mathrm{PNS}$. The total magnetic energy in the models of moderate or high resolutions (MRes, MRes10 and HRes) continues to grow rapidly. Interestingly, HRes only shows a higher total magnetic energy in the PNS after 0.20\,s, despite having a much more dramatic increase in magnetic energy in the gain region (Figure~\ref{fig:Gain}). This is likely due to differences in the nonlinear magnetic and kinetic energy dynamics inside the PNS. When comparing $E_\mathrm{kin}$, for the more strongly magnetised models, they show a similar trend to the magnetic energies. In particular, we see a general trend of higher $E_\mathrm{kin}$ with higher resolution, however, HRes shows very similar values to MRes. Surprisingly, when comparing MRes and vLRes to their counterparts with weaker initial magnetisation, MRes10 and vLRes10 show clearly lower $E_\mathrm{kin}$ in the PNS.

\begin{figure*}
\centering
    \begin{subfigure}{0.32\linewidth}
        \includegraphics[width=\linewidth]{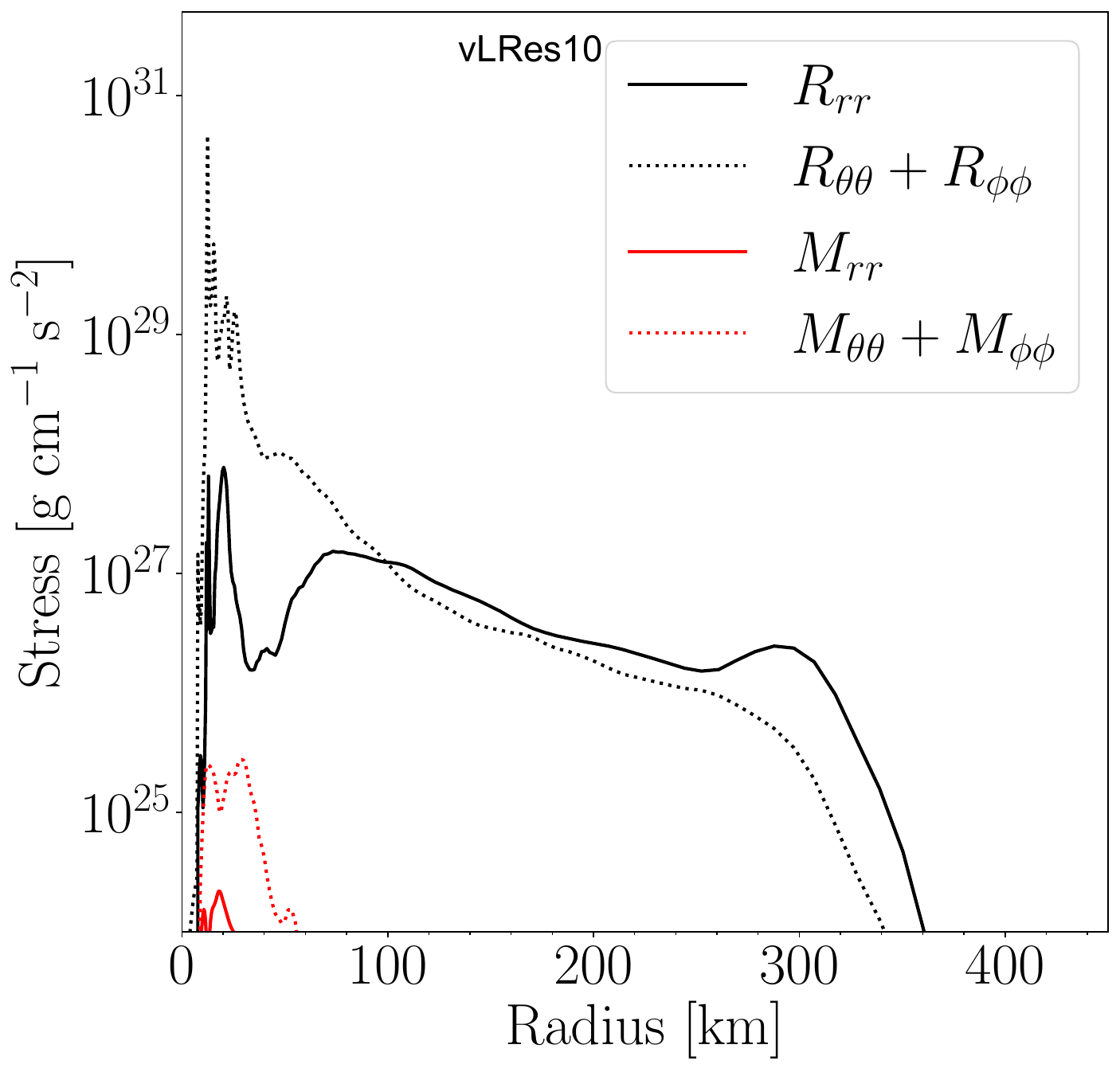}
    \end{subfigure}
\hfil
    \begin{subfigure}{0.32\linewidth}
        \includegraphics[width=\linewidth]{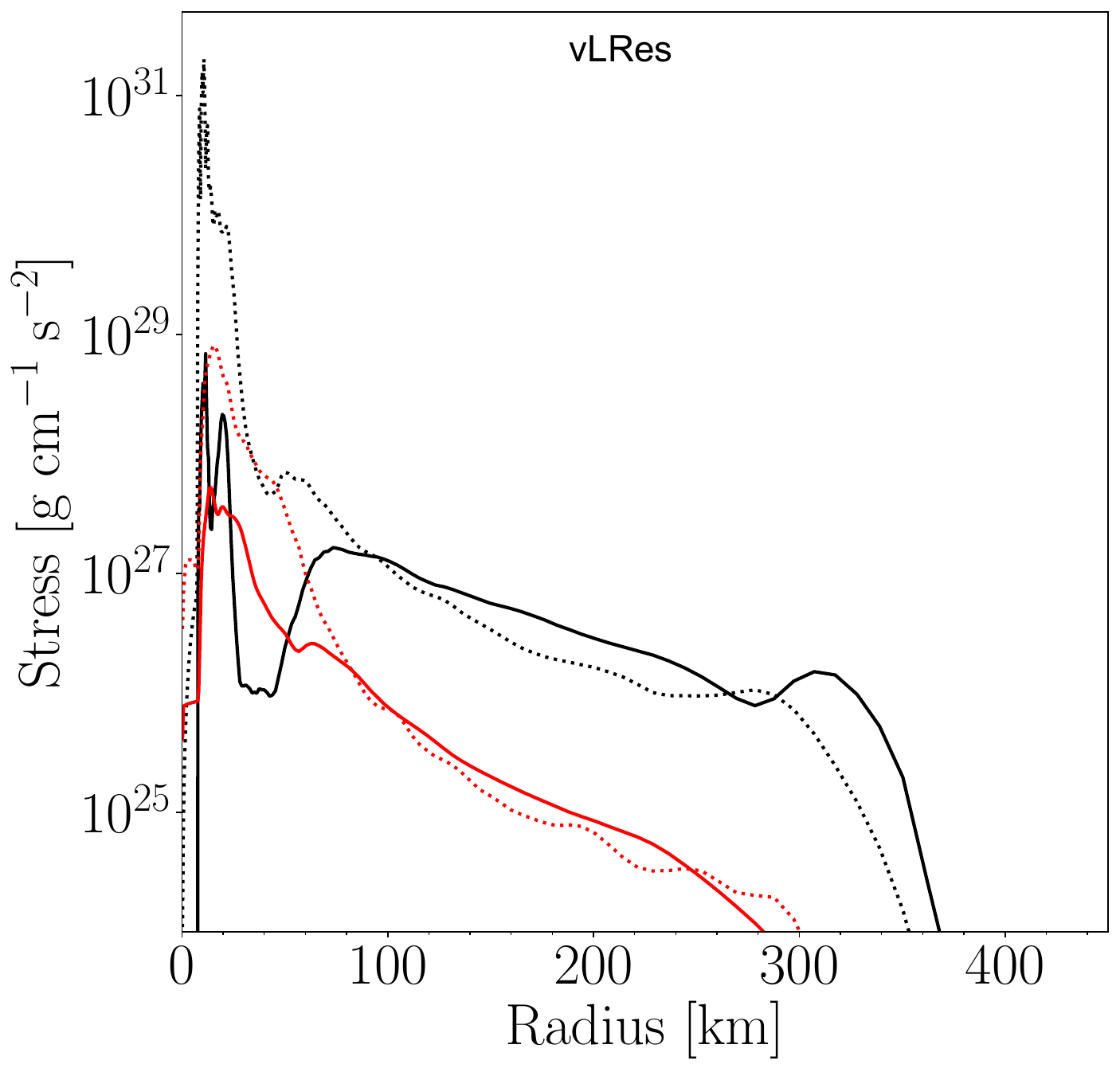}
    \end{subfigure}
\hfil
    \begin{subfigure}{0.32\linewidth}
        \includegraphics[width=\linewidth]{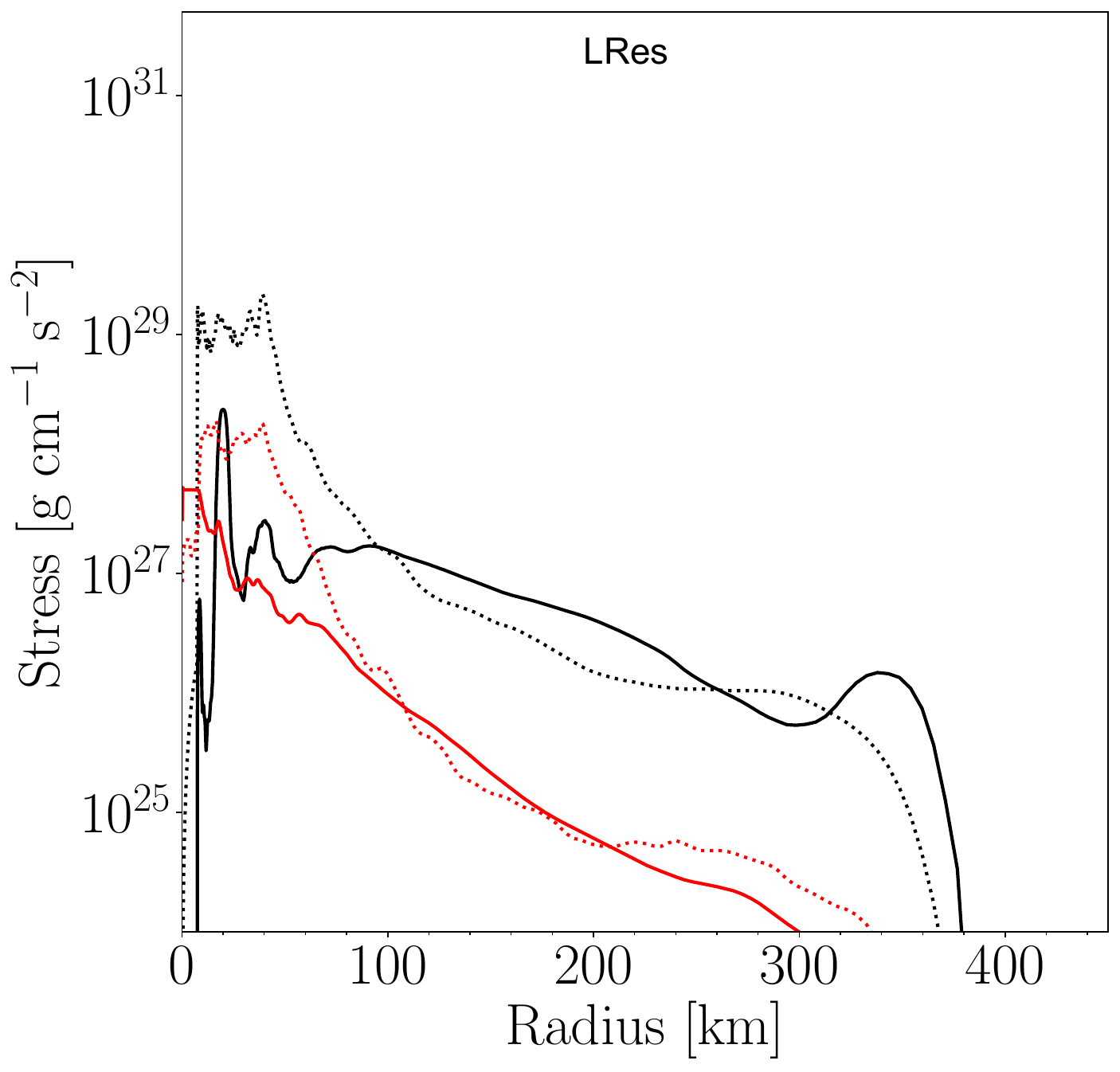}
    \end{subfigure}

    \begin{subfigure}{0.32\linewidth}
        \includegraphics[width=\linewidth]{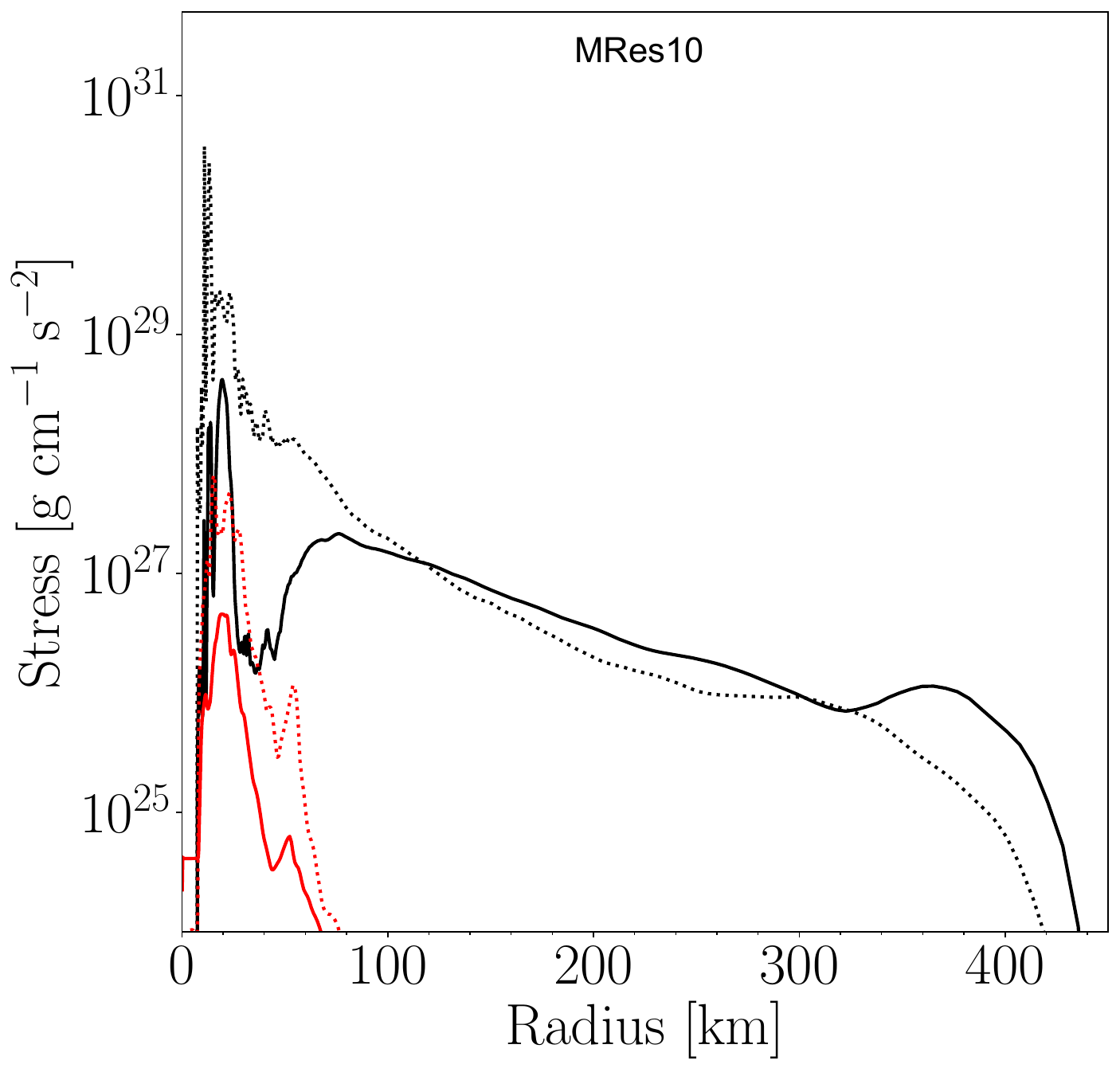}
    \end{subfigure}
\hfil
    \begin{subfigure}{0.32\linewidth}
        \includegraphics[width=\linewidth]{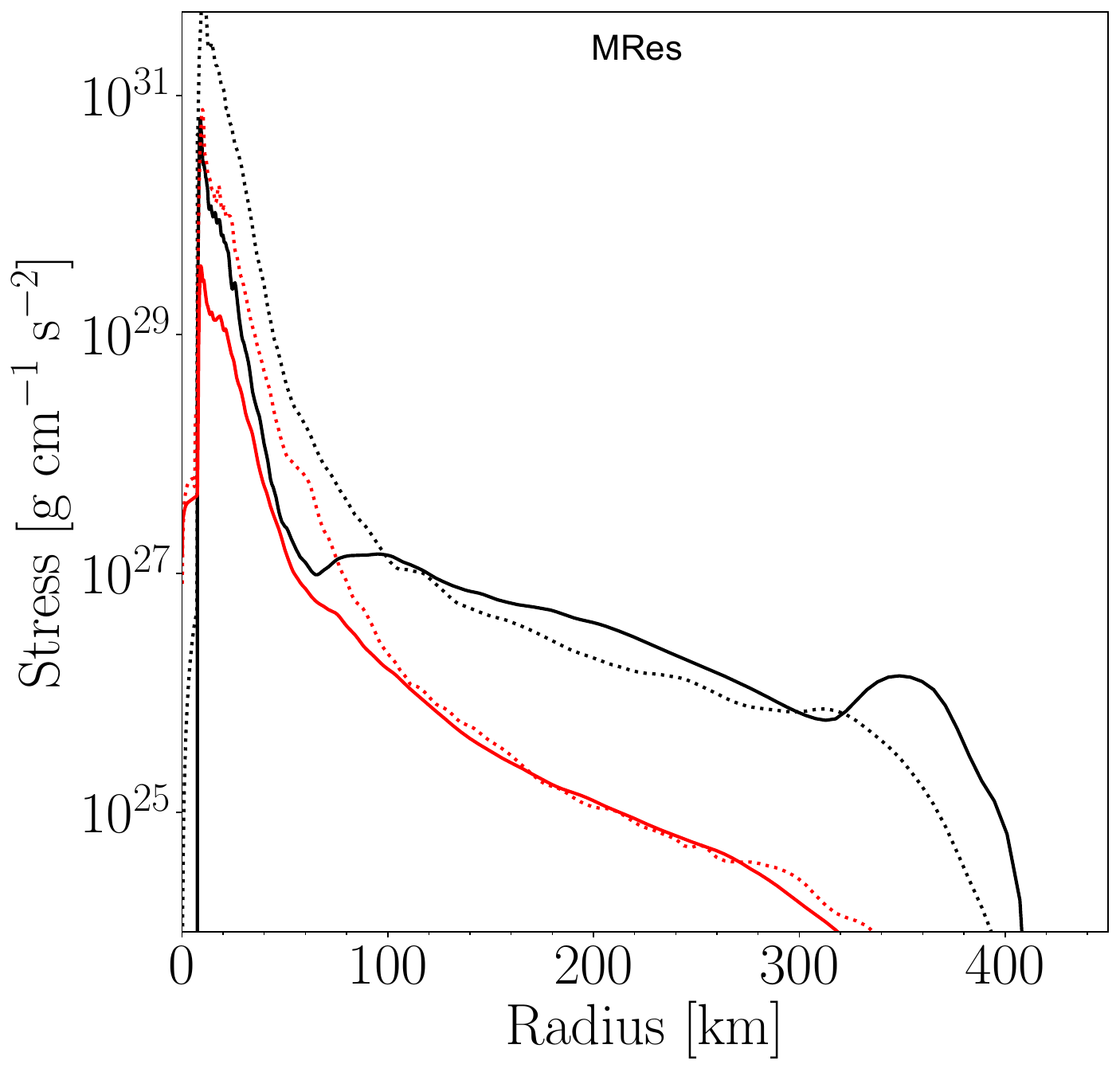}
    \end{subfigure}
\hfil
    \begin{subfigure}{0.32\linewidth}
        \includegraphics[width=\linewidth]{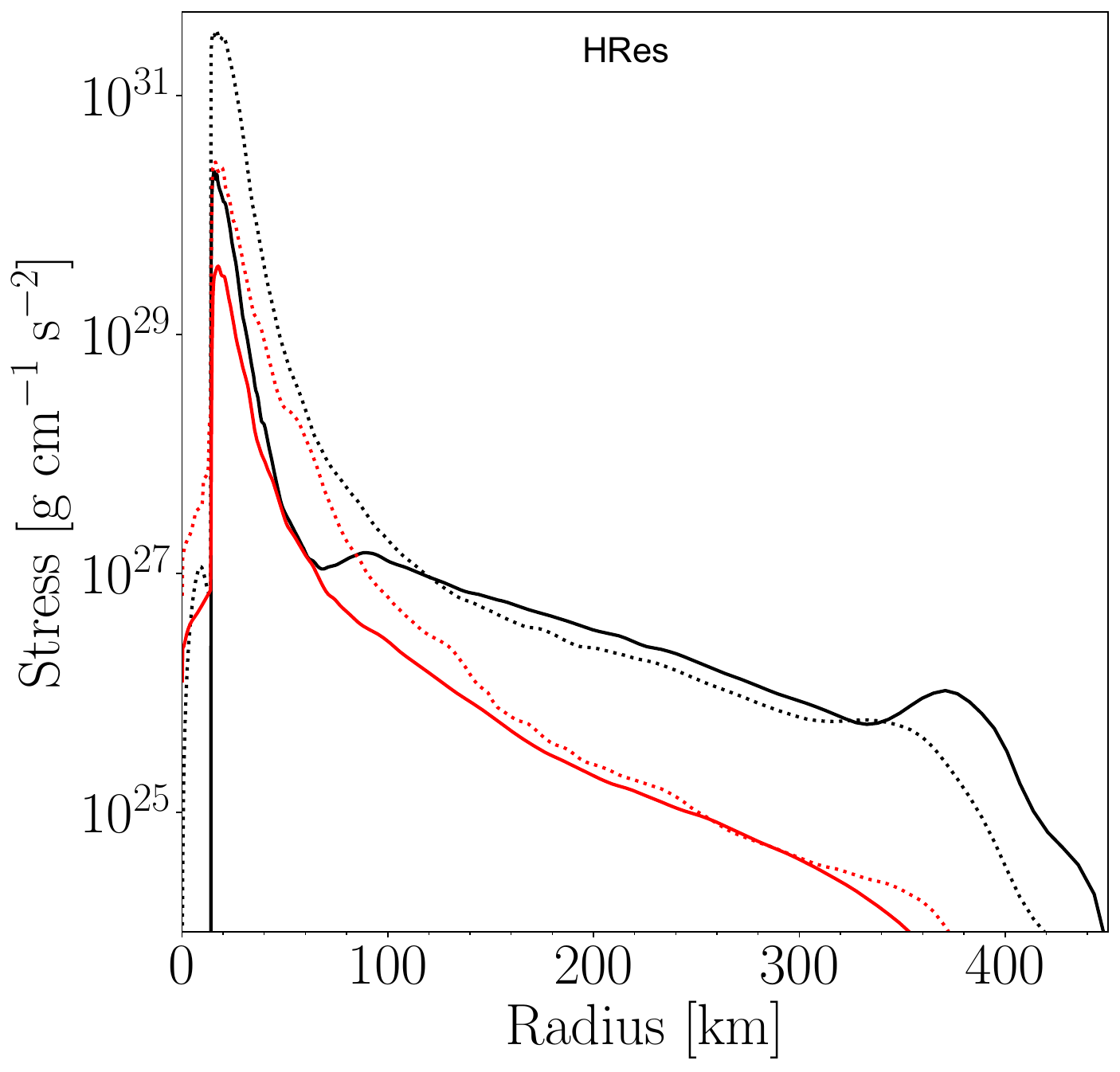}
    \end{subfigure}
\caption{Angle-averaged diagonal components of the kinetic (Reynolds, black) and magnetic (Maxwell, red) stresses for all simulated models. Solid lines depict the radial, $rr$ component, while dotted lines are the angular, $\theta\theta + \phi\phi$ components of the stress tensors, respectively.
}
    \label{fig:Stresses}
\end{figure*}

To better understand these dynamics, we compare the spherically-averaged diagonal components of the kinetic (Reynolds) and magnetic (Maxwell) stress tensors $R_{ij}$ and $M_{ij}$. $R_{ij}$ and $M_{ij}$ are computed as

\begin{eqnarray}
R_{ij}&=& \langle \rho v_i v_j\rangle, 
\label{eqn:Stress1}\\
M_{ij}&=& \frac{1}{8\pi}\langle B_i B_j\rangle,
\label{eqn:Stress2}
\end{eqnarray}
and shown for all six models at 0.20\,s in Figure~\ref{fig:Stresses} (in order left to right, top to bottom; vLRes10, vLRes, LRes, MRes10, MRes, HRes). In each plot, solid lines visualise the radial stresses, while dotted lines depict the angular stresses. As the PNS is still evolving on the timescales of our simulations, we note that the coupling between the PNS convection zone and the surrounding stable regions is not yet in equilibrium. We choose to do this comparison at 0.20\,s since PNS convection has already begun for all models at this time. These stresses are shown up to a radius of 450\,km, to include the radius of the shock at this time for all models. For all models, the radial kinetic stress ($R_{rr}$) shows an increase at the average shock radius (between 350\,km and 410\,km depending on the model). We note that even for the strongly magnetised models, at our highest resolution parameters, the magnetic stresses fall off very steeply with radius, and are several orders of magnitude weaker than the kinetic stresses at the shock interface. This is in line with what we see when comparing their energy fluxes in Figure~\ref{fig:Flux}.

Looking at our more strongly magnetised models (vLRes, LRes, MRes, HRes), we find that the magnetic stresses ($M_{ij}$) in the inner regions start to become comparable to their kinetic stress counterparts ($R_{ij}$). In particular, we see that the angular magnetic stresses, $M_{\theta\theta} + M_{\phi\phi}$, are comparable to $R_{rr}$. We find that the strongly magnetised cores show an extended PNS convection zone that appears to fill the entirety of the PNS that is resolved in 3D, barring the innermost ~10\,km of our grid that is treated in spherical symmetry.
This change in stress profiles can be seen most clearly when comparing MRes10 to MRes. We suspect that we do not see an equivalent change when comparing vLRes10 to vLRes due to the higher numerical dissipation in these models.

We note that while this extended PNS convection zone does not appear in LRes (Figure~\ref{fig:Stresses}), it is in a transitional phase at the snapshot we have chosen. We find that the multi-peaked structure of $R_{rr}$ in LRes merges into a one extended PNS convection region by 0.30\,s, similar to what we see in the higher resolution models, MRes and HRes. Similarly, the angular stress component, ${R_{\theta\theta}} + {R_{\phi\phi}}$ for LRes is also in transition, and soon settles to a single sharp peak at the inner boundary similar to MRes and HRes, albeit at a lower magnitude.

\begin{figure*}
\centering
    \begin{subfigure}{0.45\linewidth}
        \includegraphics[width=\linewidth]{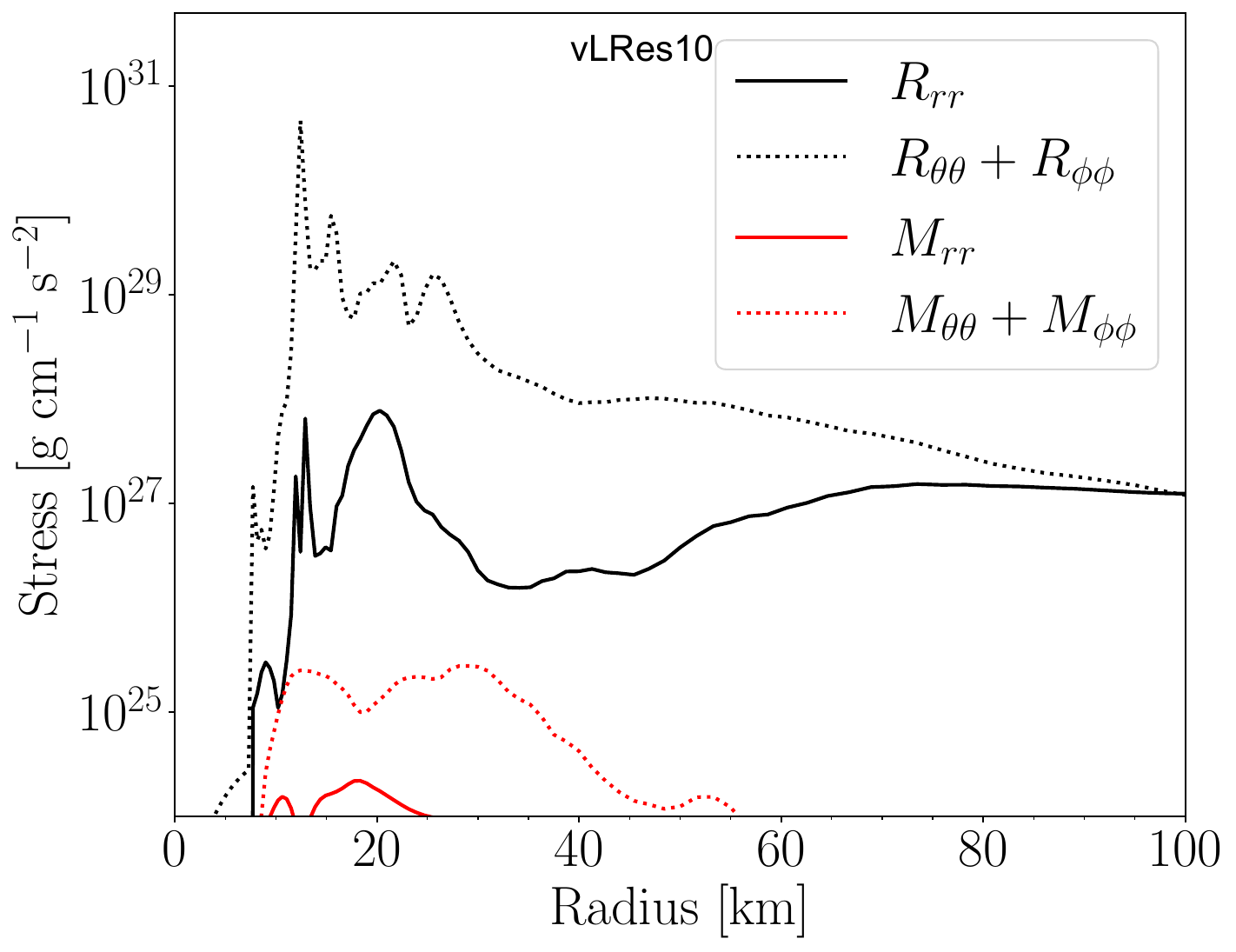}
    \end{subfigure}
    \begin{subfigure}{0.45\linewidth}
        \includegraphics[width=\linewidth]{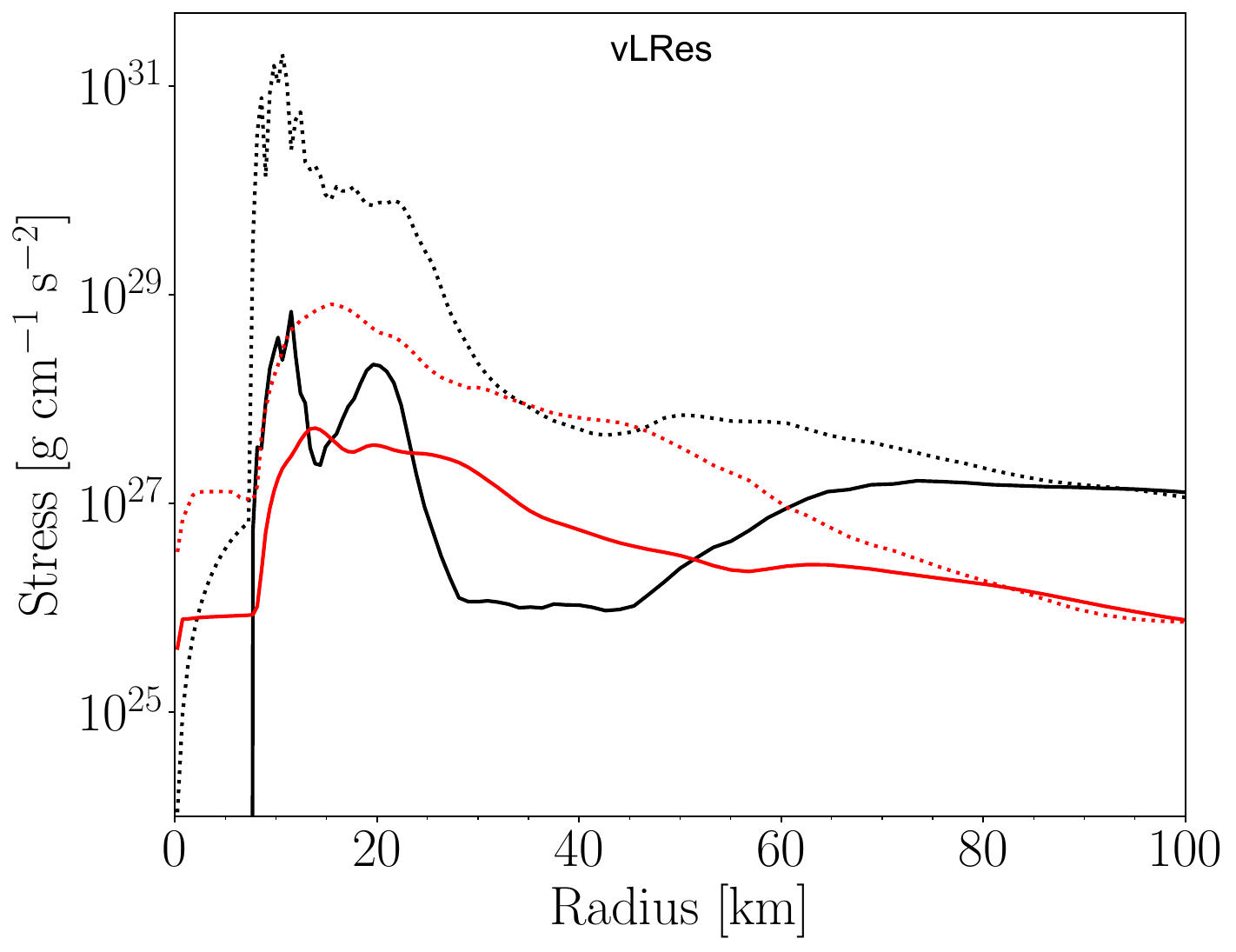}
    \end{subfigure}

    \begin{subfigure}{0.45\linewidth}
        \includegraphics[width=\linewidth]{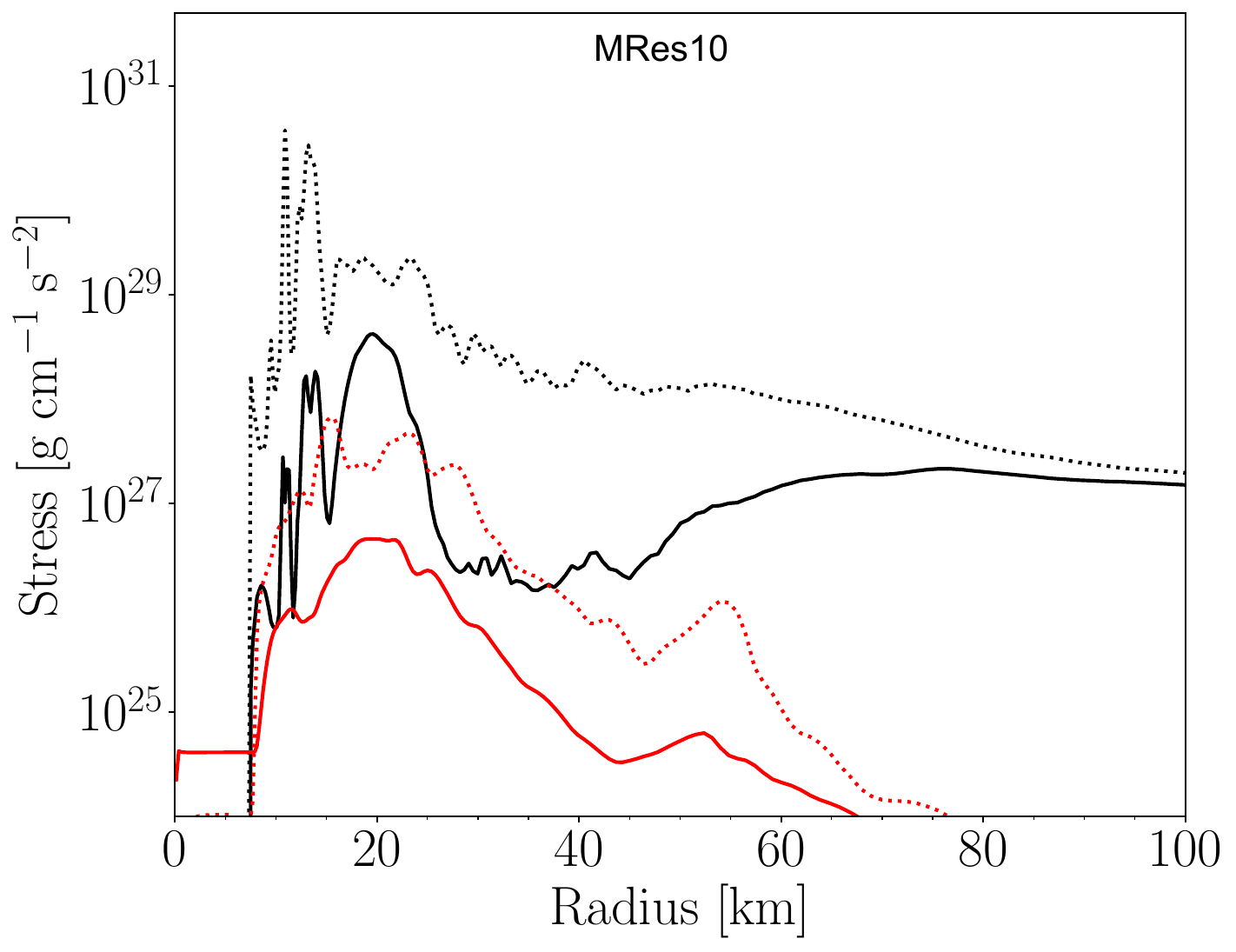}
    \end{subfigure}
    \begin{subfigure}{0.45\linewidth}
        \includegraphics[width=\linewidth]{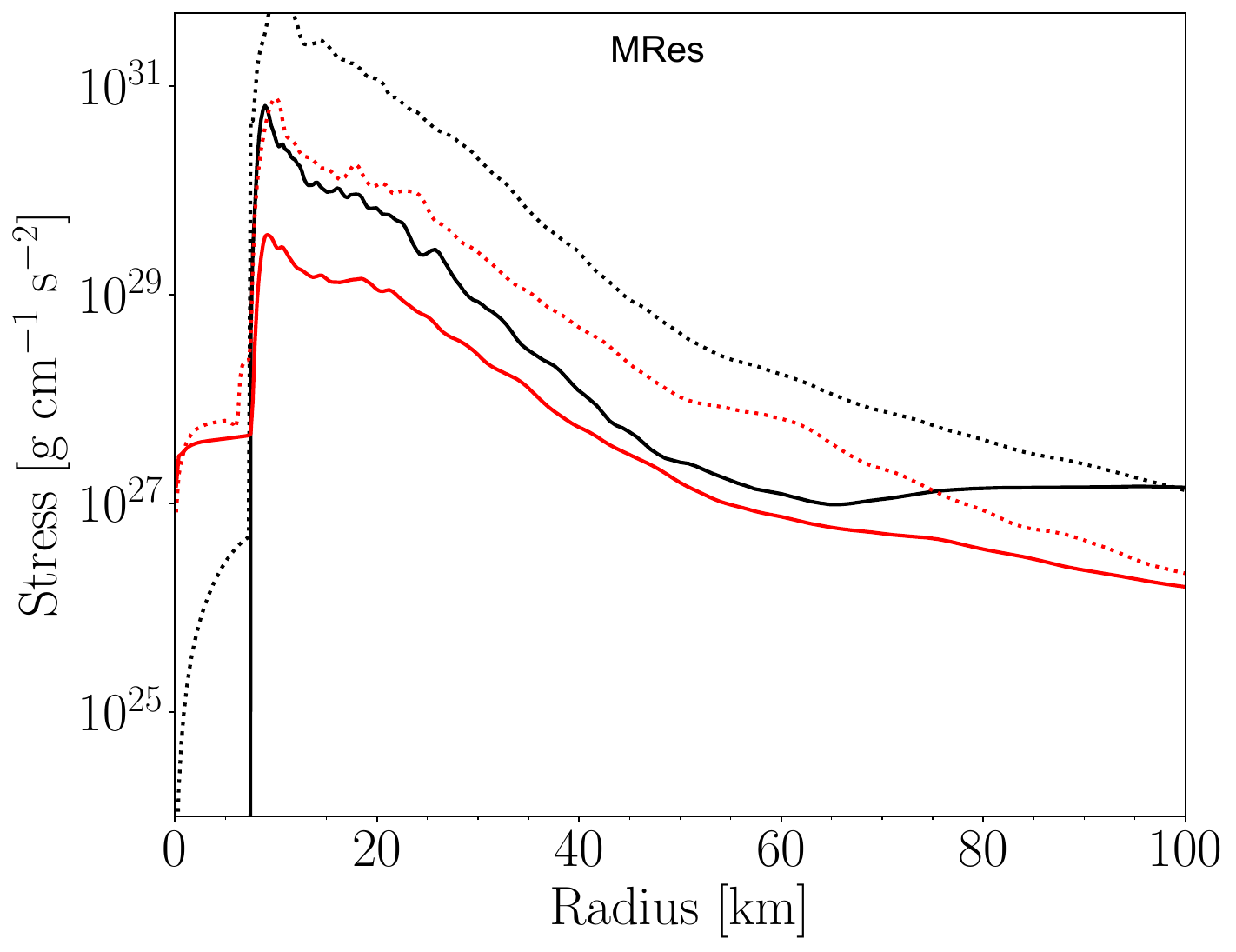}
    \end{subfigure}

\caption{Angle-averaged diagonal components of the kinetic (Reynolds, black) and magnetic (Maxwell, red) stresses for a subset of simulated models in the innermost 100\,km. Solid lines depict the radial, $rr$ component, while dotted lines are the angular, $\theta\theta + \phi\phi$ components of the stress tensors, respectively, for models vLRes10, vLRes, MRes10 and MRes.
}
    \label{fig:Stresses_PNS}
\end{figure*}

For a clearer comparison of the stresses at different initial magnetisations, we show the stresses in the PNS (the inner 100\,km) of these vLRes and MRes, along with their more weakly magnetised counterparts, vLRes10 and MRes10, in Figure~\ref{fig:Stresses_PNS}.

For the two weakly magnetised models, vLRes10 (top left) and MRes10 (bottom left), the magnetic stresses are far below equipartition with the kinetic stresses and do not impact the dynamics inside the PNS. As we have already discussed, the higher resolution of MRes10 allows for a higher rate of magnetic field amplification compared to vLRes10, however, it is still below being dynamically relevant at this time. Given that PNS convection has begun at around 15-30\,km (depicted by the spike in radial kinetic stress), it is unclear if this will continue to amplify the magnetic field strength. 

We attribute these differences in the PNS stresses to a complex exchange between magnetic energy, turbulent kinetic energy and rotational energy within the forming PNS. We find that both rotational and turbulent energy are converted into magnetic energy, which is then redistributed. This phenomenon is non-linear, since a stronger magnetic field also leads to more efficient spin-up of the PNS, and hence an increase in rotational energy. We also see a consequence of this in the kinetic energy $E_\mathrm{kin}$ of motions in the PNS in Figure~\ref{fig:Mag}. Here, for models with the same spatial resolution (e.g. MRes and Mres10), the more strongly magnetised model has systematically higher values of $E_\mathrm{kin}$. These non-linearities are coupled with the fact that the PNS is still contracting, making an accurate tracking of the energy transfers difficult.
The models that are weakly magnetised, or equivalently, numerically diffusive, like vLRes, do not show similar structures. As the magnetic field strengths are still growing in the PNS of MRes, it is unclear if this will eventually lead to a similar effect, and qualitatively change in its stresses to more closely resemble that of MRes. As the inner $\approx$10\,km of our domain are treated in spherical symmetry, we see a sharp, artificial rise in all stress components at this interface (right panels of Figure~\ref{fig:Stresses_PNS}). We note that this effect may be somewhat exaggerated at the radial inner boundaries of our models. This is particularly noteworthy as the PNS is being spun up, leading to substantial differential rotation at this interface, which further amplifies the magnetic fields.

Although MRes and HRes have qualitatively similar stress distributions, we see that the peak values of both the magnetic and kinetic stresses are slightly higher in MRes. In particular, we see that the angular components of the magnetic and kinetic stresses for MRes are higher than those for HRes. We attribute this to the higher stochastic spin-up rate in MRes, which then injects more rotational energy into other components of $M_{{ij}}$ and $R_{{ij}}$.

\begin{figure*}
\centering
    \subfloat{\includegraphics[width=0.48\linewidth]{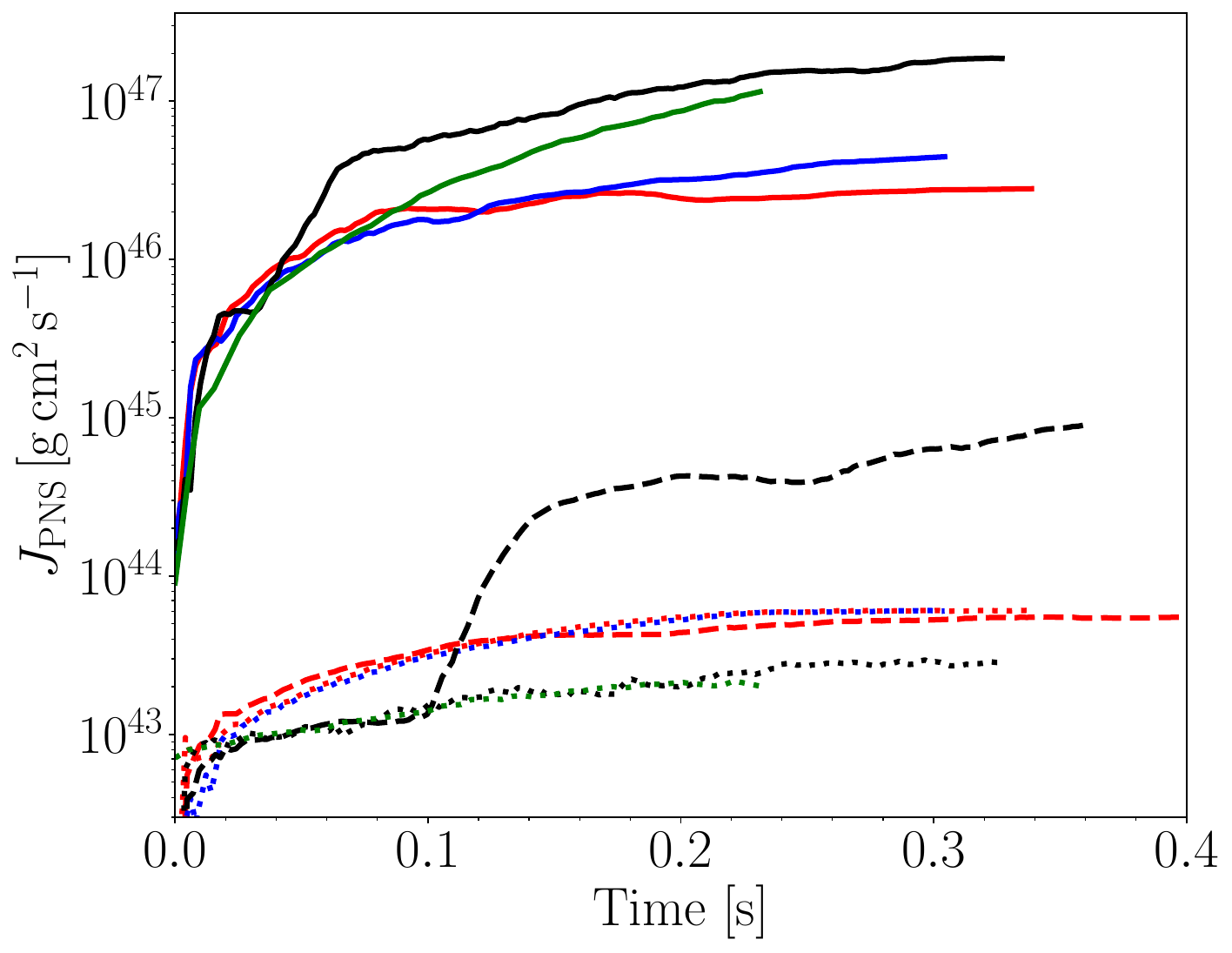}}
\hfil
    \subfloat{\includegraphics[width=0.48\linewidth]{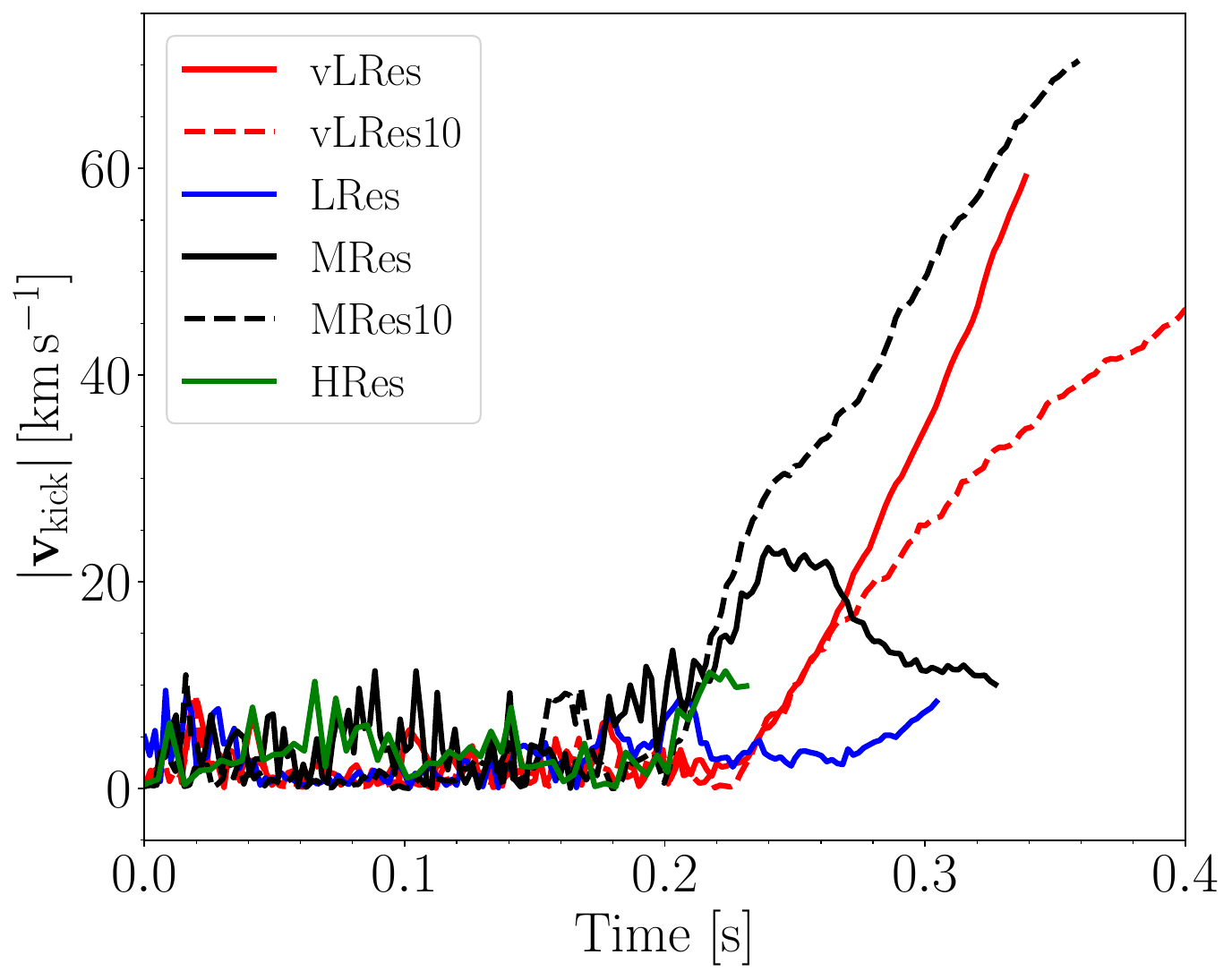}}
\caption{\normalsize Evolution of the PNS total angular momentum, ${J_\mathrm{PNS}}$, (left) and estimated PNS kick, $|\mathbf{v_\mathrm{kick}}|$, (right) for all models. Dotted lines in the ${J_\mathrm{PNS}}$ plot shows the purely hydrodynamic component of the angular momentum for each of the strongly magnetised models (vLRes, LRes, MRes and HRes).}
    \label{fig:NS_spin_kick}
\end{figure*}

We plot the total angular momentum in the PNS of all our models in Figure~\ref{fig:NS_spin_kick} (left). The angular momentum $\mathbf{J}$ of the PNS, is obtained by integrating the magnetic torques and hydrodynamic angular momentum fluxes onto the PNS in time and over the surface of a sphere of radius $30\, \mathrm{km}$,
\begin{equation}
    \frac{\ud \mathbf{J}}{\ud t}
    =-\oint (\mathbf{r}\times  \mathbf{v}) \rho v_r\,\ud A
    +\frac{1}{4\pi}\oint (\mathbf{r}\times \mathbf{B}) B_r\,\ud A.
\end{equation}

Dashed lines depict our models with initially weaker magnetisation, and dotted lines are just the hydrodynamic angular momentum terms for models with initial $B_\mathrm{pol,tor} = 10^{12}\,\mathrm{G}$ as described in Table~\ref{tab:resolution} (i.e. vLRes, LRes, MRes and HRes). 

Comparing the total angular momenta of the strongly magnetised models to the hydrodynamic contributions, we see that the magnetic torques are by far the dominant term that leads to the spin-up of the neutron star. This becomes particularly evident when we compare just the hydrodynamic contributions of the angular momentum to the total angular momenta of the weakly magnetised models. Starting with the very low resolution models, we see that the total angular momentum for vLRes10 is very similar to just the hydrodynamic component of vLRes, sitting at around $\approx4\times10^{43}\mathrm{g\,cm^2\,s^{-1}}$. Including the magnetic torques, however, vLRes increases its angular momentum by several orders of magnitude to $\approx2\times10^{46}\mathrm{g\,cm^2\,s^{-1}}$. Similarly, the hydrodynamic component of MRes and the total angular momentum of MRes10 initially lie on top of one another at $\approx9\times10^{42}\mathrm{g\,cm^2\,s^{-1}}$. Unlike the very low-resolution models, at these higher resolutions, the magnetic energy of MRes10 continues to grow both in the gain region (Figure~\ref{fig:Gain}) and in the PNS (Figure~\ref{fig:Mag}). This growth eventually leads to the PNS of MRes10 spinning up due to a contribution from magnetic torques. A similar angular momentum evolution can be seen in the s24 model of \citet{Sykes2024}, where magnetic torques start playing an important role only at late times, once the magnetic fields start to be amplified substantially in the gain region. By the end of our simulation, the integrated angular momentum of MRes10 is at $\approx10^{45}\mathrm{g\,cm^2\,s^{-1}}$ and is still increasing. 

We note that despite its higher resolution, at the end of our simulation, the angular momenta of HRes os lower than that of MRes. This is likely due to stochastic differences in the initial explosion and the corresponding mass accretion. Comparing their trends at the end of our simulations in Figure~\ref{fig:NS_spin_kick} (left), the angular momenta of both PNS are growing, with HRes growing at a faster rate than MRes. So, it is unclear what their final values will be.

Using the analytic approximation for the neutron star moment of inertia from \citet{Lattimer2005},
\begin{equation}
    I \approx 0.237 M_\mathrm{grav}R^2
    \left[
    1 + 4.2\left(\frac{M_\mathrm{grav}}{M_{\odot}}\frac{\mathrm{km}}{R}\right)
    + 90\left(\frac{M_\mathrm{grav}}{M_{\odot}}\frac{\mathrm{km}}{R}\right)^4
    \right],
\end{equation}

and 
\begin{equation}
    M_\mathrm{by} = M_\mathrm{grav} + 0.084M_{\odot}(M_\mathrm{grav}/M_{\odot})^2,
\end{equation}

the angular momentum can be translated into a neutron star spin period $P=2\pi / (J/I)$. 
Assuming a final neutron star radius $R$ of $12\, \mathrm{km}$, we estimate the moment of inertia to be $\mathrm{\approx1.52 \times 10^{45}g\,cm^2}$, which corresponds to spin periods of vLRes: 0.178\,s vLRes10: 280\,s, LRes: 0.114\,s MRes: 0.024\,s, MRes10: 8.12\,s, HRes: 0.073\,s. 

We next consider the kick velocity of the newly born neutron star, which we plot in Figure~\ref{fig:NS_spin_kick} (right). Following \citet{Scheck2008}, we evaluate the kick indirectly from the momentum of the ejecta assuming angular momentum conservation since the innermost part of our grid is treated in spherical symmetry and does not allow the PNS to move freely. 

We estimate the kick velocity as,
\begin{equation}
    \mathbf{v}_\mathrm{kick} = - \frac{1}{M} \int_{r>R_\mathrm{gain}} \rho \mathbf{v} \,\ud V,
\end{equation}

where $M$ is the total gravitational mass, although, similarly to our calculation of the angular momentum above, the difference between the baryonic and gravitational masses in these cases is small. 
Unsurprisingly, due to the weak explosion energy of this progenitor, we find relatively weak kick velocities for all our models. By the end of our simulation runs, the PNS of MRes10 has the highest kick velocity at 71\,km/s. Although MRes has the more powerful explosion, Figure~\ref{fig:2D_morph} shows that MRes10 develops more asymmetric flow morphology. In MRes, the outflow appears very spherically symmetric, so that the momentum carried by the ejecta is more nearly balanced across opposite hemispheres. By contrast, MRes10 exhibits a more one-sided large-scale outflow structure at these early times, leading to less cancellation of the ejecta momentum and therefore a larger net kick of the proto-neutron star. This comparison emphasises that the kick is governed not simply by the overall explosion strength, but by the global asymmetry of the explosion. We note that in our set of models, the morphology of the explosion and the related PNS kick is a stochastic process, as we see no correlation to either resolution or magnetisation.

The PNS kicks for our models are on the low end compared to pulsar observations \citep{hobbs_05}, however, the $E_\mathrm{expl}$ and $\mathbf{v}_\mathrm{kick}$ for all models are still growing. As has been shown by long-timescale CCSN simulations \citep{Burrows2024, Janka2025}, we require much longer simulations to make any conclusions on the final neutron star kicks and spin.

\section{Conclusions}
\label{sec:conclusion} 

We conducted the first resolution study of 3D MHD simulations of neutrino-driven supernovae, using the CoCoNuT-FMT code. We simulated a set of four resolutions, with two different initial dipole magnetic fields ($10^{12}\, \mathrm{G}$ and $10^{10}\, \mathrm{G}$)  for a non-rotating $13\, \mathrm{M_{\odot}}$ progenitor and evaluate their effect on shock revival, explosion dynamics, and the properties of the compact remnant. The list of models and initial conditions are summarised in Table \ref{tab:resolution}.

Our simulations show a correlation between resolution and the bulk dynamics of the explosion, with better resolved models having higher diagnostic explosion energies and shock radii. This is similar to what has been found in resolution studies of purely hydrodynamic CCSN \citep[e.g.,][]{Nagakura2019, Melson2020}, however, in our models, this effect is not driven by additional turbulent energy. Turbulent energy is instead efficiently converted into magnetic energy, increasing the contribution of magnetic fluxes that aid in driving the shock. Aside from the very low resolution models, all others achieve shock revival simultaneously, regardless of resolution and magnetisation, however, energetics differ at later times. We find that the strong magnetic fields do not drive shock revival itself; however, the strongly magnetised, higher resolution models maintain higher magnetic energies, which contribute to the explosion energy.

For our grid of magnetic CCSN, we find that neutrino luminosity and mean energy are affected by our choice of grid resolution. At higher resolution, neutrino luminosity and mean energy decrease. The strong magnetic fields distort the PNS, causing it to be more extended towards the poles, changing the locations of part of the neutrino emissions. The higher-resolution models are more turbulent and less diffusive, so they can sustain a higher magnetic field strength, and hence, an increased PNS distortion. We thus see a trend in neutrino properties. 
These difference in neutrino luminosity leads to differences in neutrino interaction rates, leading to changes in $Y_\mathrm{e}$ distribution in the PNS surface region.

At higher resolutions, we find that magnetic field amplification continues in the weakly magnetised progenitors (MRes10). These models exploded quickly without the need for additional magnetic fluxes to drive the explosion. However, it is possible that for a model that experiences longer stalled shocks, with enough time, even a weakly magnetised model could amplify its magnetic field to have a dynamic impact on the supernova. 

Although these simulations have not reached saturation of their properties, we estimate the spin and kick imparted on their forming PNS. We find PNS kicks in the range of 10\,km/s to 71\,km/s, however, these values are still growing. When strong magnetic fields are present, magnetic torques dominate the spin-up of the neutron star, leading to significantly faster spin periods for the initially strongly magnetised models. We find spin periods between 0.024\,s and 0.178\,s for strongly magnetised models. Once the magnetic field strengths have amplified enough, the magnetic torques also play a significant role in the spin-up of the well-resolved, initially weakly magnetised model (MRes10). MRes10 ends with a spin period of 8.12\,s, as opposed to 280\,s for vLRes10, which never experiences significant spin-up by magnetic torques.

We find that the structure and strength of the PNS convection that develops are dictated by both the strength of the magnetic field and resolution (which enables stronger fields to be sustained). The resolution, magnetic energy and rotational energy of the different models lead to a complex energy exchange within the PNS. This leads to a more extended region of PNS convection, with much higher stresses for models that maintain stronger magnetic fields and rotation. A much longer simulation of this model would be necessary to determine how this change would impact the neutrino-driven wind and the development of the magnetic fields on the PNS at late times. All initially strongly magnetised models have magnetar field strengths by the end of our simulations. The magnetic field strength at the PNS surface is correlated with its initial conditions. However, this field may still be amplified, so even initially weakly magnetised models may attain magnetar field strengths, as seen in \citet{Varma2023}.

Our study suggests that the details of MHD supernova models are sensitive to the spatial resolution. Due to the lower numerical diffusivity at higher resolutions, stronger magnetic energies can develop more rapidly both in the gain region and in the PNS, impacting their evolution. Unlike in purely hydrodynamic CCSN \citep{Melson2020}, where saturation is found at resolutions of $1^\circ$, even at $0.7^\circ$ in our MHD simulations, we have not achieved saturation of bulk properties in these supernovae. This is particularly evident in the difference in magnetic energy and magnetic-to-kinetic energy ratios in the gain region. We show that under-resolving magnetic models can lead to quantitatively different results when strong magnetic fields cannot be maintained due to high numerical diffusivity. We stress the need to consider the role of magnetic fields in supernova explosions more generally, with more realistic initial magnetic field strengths and geometries, and multidimensional progenitors to understand the explosion dynamics and the progenitor-remnant connection fully. To understand and quantify the added uncertainties in these MHD simulations, we require both long-time resolution studies as well as detailed code comparisons across independent modelling groups.

\section*{Acknowledgements}

BM acknowledges support from the ARC through Discovery Project DP240101786. VV acknowledges support from the Royal Astronomical Society Research Fellowship. This work used the DiRAC Memory Intensive service (Cosma8) at Durham University, managed by the Institute for Computational Cosmology on behalf of the STFC DiRAC HPC Facility (www.dirac.ac.uk). The DiRAC service at Durham was funded by BEIS, UKRI and STFC capital funding, Durham University and STFC operations grants. DiRAC is part of the UKRI Digital Research Infrastructure. 


\section*{Data Availability}
The data underlying this article will be shared on reasonable request to the  authors, subject to considerations of intellectual property law.



\bibliographystyle{mnras}
\bibliography{paper} 








\bsp	
\label{lastpage}
\end{document}